\documentclass[a4paper,12pt,oneside,reqno]{amsart}
 
\usepackage[utf8]{inputenc}
\usepackage{csquotes}
\usepackage{amsmath,amsfonts,amssymb}
\usepackage{tikz}
\usepackage{ dsfont }
\usepackage[headinclude,DIV13]{typearea}
\usepackage[square,sort,comma,numbers]{natbib}
\usepackage{verbatim}
\usetikzlibrary{arrows,shapes,backgrounds}
\usepackage{amssymb}
\usepackage{bm}
\usepackage{multirow}

\usepackage{float}

\DeclareMathOperator*{\minimum}{min}

\DeclareMathOperator*{\argmaxA}{arg\,max}

\graphicspath{{CTRL/},{ACT/},{ACTRE/},{OPT_ACT/},{OPT_ACTRE/},{_fig/}}

\usepackage{subcaption}
\makeatletter
\newcommand{\addresseshere}{%
  \enddoc@text\let\enddoc@text\relax
}
\makeatother

\numberwithin{equation}{section}

\begin{document}
	\title[Parameter estimation in a stochastic model for immunotherapy]{Parameter estimation and treatment optimization in a stochastic model for immunotherapy of cancer}

\author{Modibo Diabate}
\address{Modibo Diabate,
	Laboratoire Jean Kuntzmann, Univ.\ Grenoble Alpes
	F-38000 Grenoble, France}
\email{modibo.diabate@univ-grenoble-alpes.fr}

\author{Loren Coquille}
\address{Loren Coquille,
	Univ.\ Grenoble Alpes, CNRS, Institut Fourier, F-38000 Grenoble, France}
\email{loren.coquille@univ-grenoble-alpes.fr}

\author{Adeline Samson}
\address{Adeline Samson,
	Laboratoire Jean Kuntzmann, Univ.\ Grenoble Alpes
	F-38000 Grenoble, France}
\email{adeline.leclercq-samson@univ-grenoble-alpes.fr}


\thanks{This work has been supported by the {\color{black}Malian Government} and the LabEx PERSYVAL-Lab (ANR-11-61 LABX-0025-01). 
	The authors would like to warmly thank Meri Rogava, Thomas Tüting and Michael Hölzel for fruitful discussions and for providing important medical informations, Jennifer Landsberg for providing the database used in this paper,
	and the Data Institute Univ.\ Grenoble Alpes for allowing computing time on the cluster "Luke"
}

\maketitle

{\bf Abstract.} 
Adoptive Cell Transfer therapy of cancer is currently in full development and mathematical modeling is playing a critical role in this {\color{black}area. We study} a stochastic model developed by Baar et al.\ \cite{Loren} for modeling immunotherapy against melanoma skin cancer. 
First, we estimate the parameters of the deterministic limit of the model based on biological data of tumor growth in mice.
A Nonlinear Mixed Effects Model is estimated by the Stochastic Approximation Expectation Maximization algorithm. 
With the estimated parameters, we head back to the stochastic model and calculate the probability that the T cells all get exhausted during the treatment. We show that for some relevant parameter values, an early relapse is due to stochastic fluctuations (complete T cells exhaustion) with a non negligible probability. Then, focusing on the relapse related to the T cell exhaustion, we propose to optimize the treatment plan (treatment doses and restimulation times) by minimizing the T cell exhaustion probability in the parameter estimation ranges.

\bigskip


\section{Introduction}
Cancer is a group of more than 100 different diseases causing a large number of deaths a year worldwide \cite{ferlay2015cancer}. Cells start   growing uncontrollably due to genetic changes which impair their normal evolution. It can develop almost anywhere in the body. Cancer is a complex disease, difficult to study biologically (expensive and time consuming to experiment with animals and humans). In this context, mathematical modeling can be an excellent tool for emitting or confirming biological assumptions with less expensive experiments.  

When it is diagnosed quickly, cancer can be treated by chemotherapy, surgery, radiotherapy or by immunotherapy \cite{airley2009cancer, sabel2006principles, hawley2013principles, liebelt2016principles}. Immunotherapy is a recent treatment that {\color{black}activates} the immune system to kill cancer cells.
In this paper we are interested by the Adoptive Cell Transfer (ACT) therapy to treat melanoma in mice with cytotoxic T cells, 
as experimented  in 2012 by Landsberg et al.\ \cite{Landsberg}. This immunotherapeutic approach involves the stimulation of T cells which recognize one specific type of melanoma tumor cells (differentiated melanoma cells) through the antigen gp100 on their surface. The stimulated T cells are then able to kill these differentiated melanoma cells. 

\begin{figure}
\begin{center}
  \includegraphics[width=0.87\linewidth, height=0.22\linewidth]{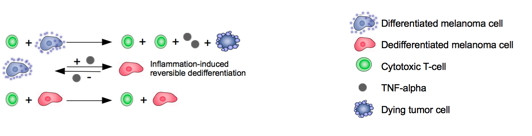}
  \caption{\scriptsize{Cell interactions in ACT therapy: T cells kill  differentiated melanoma  cells and  produce  T cells and cytokines $TNF_{\alpha}$ (line 1);  switch between differentiated and dedifferentiated melanoma cells is reversible (line 2); T cells cannot kill  dedifferentiated melanoma  cells (line 3).}} 
  \label{Cellules_Immunotherapy}
  \end{center}
\end{figure}

	The authors of \cite{Landsberg} showed that during the inflammation induced by the therapy, pro-inflammatory cytokines called $TNF_{\alpha}$ (Tumor Necrosis Factor) which are released in the body enhance a  cell-type switch: the markers on the differentiated cancer cells disappear. They become dedifferentiated, and cannot get killed by T cells anymore.  The resulting tumor tissue consists of both differentiated and dedifferentiated melanoma cells. Note that the switch is reversible (i.e. the melanoma cells can recover their initial  type) and it does not require cell division or mutation. 
Figure \ref{Cellules_Immunotherapy} illustrates the described interactions between cells during the treatment.

In immunotherapy as in other cancer treatments,  relapse is one of the main problems. The authors of  \cite{Landsberg, Loren} describe two kinds of relapse in the ACT therapy. 
First, T cells only recognize the differentiated cancer cells (through the antigen gp100) and not the dedifferentiated cancer cells which they are thus not capable to kill. The growth of dedifferentiated cells  induces a relapse.
Second, T cells can become exhausted and no longer able to kill differentiated cancer cells thus causing a relapse. This problem was addressed in \cite{Landsberg} by T cells restimulation, which only lead to a delay in the occurrence of the relapse. 


In these recent years, a lot of potentials have been seen in immunotherapy treatments, including the possibility of higher effectiveness \cite{sagiv2018eradication} with lower side effects. But these promising new treatments are still to be understood, and
long-term effects have not yet been studied in clinical trials, which are long and costly. 
 An alternative is to study treatment failure and relapse through the development of mathematical models. 
Numerous deterministic models, mainly based on partial differential equations or ordinary differential equations, were developed for cancer study \cite{novozhilov2006mathematical,sun2005deterministic,watanabe2016mathematical,barbolosi2009mathematical}. Stochastic counterparts include branching processes, individual-based or diffusion models \cite{Durrett15,cattiaux2016stochastic,ochab2005pattern,Loren} and often a combination of both \cite{quaranta2005mathematical,anderson1998continuous}. The parameter estimation of those models often involve Mixed Effects Models \cite{pinero2000mixed} which are well adapted to observed longitudinal data \cite{desmee2015nonlinear,brant2003screening,honerlaw2016biopsychosocial}. 

{The purpose of this paper is to provide a quantitative study of the relapse and failure of skin cancer immunotherapy in mice using a stochastic individual-based model introduced in Baar et al.\ \cite{Loren} {\color{black}and real biological data provided in Landsberg et al.\  \cite{Landsberg}}. We first perform a statistical estimation of the parameters of the model, and then study the treatment relapse and failure focusing on T cell exhaustion. Finally we propose an optimization of the treatment plan.}

{\color{black}The stochastic individual-based {\color{black}model} recently proposed by \cite{Loren} aims to describe the ACT therapy {\color{black}presented in the article} \cite{Landsberg}}. {Its main features are: modeling the proliferation and the death of cancer cells and active T cells with birth and death processes \cite{kendall1948generalized} (the division of an active T cell is modeled as a birth event, and its exhaustion as a death event);} 
modeling the switch between the two types of melanoma cancer cells; taking into account the interactions between cancer cells and T cells in a predator-prey framework (see related models in \cite{berryman1992orgins, costa2016stochastic}). The different events (division, death and switch events) occur at exponential random times whose rates constitute the parameters of the model. The large population limit of the stochastic model is a deterministic differential system. 

{\color{black}The authors of \cite{Loren} have provided a set of biologically relevant parameters for which the stochastic system exhibits exhaustion of the T cells with high probability. However, these parameters have been calibrated numerically but not estimated from real data. Our first objective is thus to estimate these parameters from {\color{black}the biological data given in \cite{Landsberg}} and to provide   intervals of values defined by extracting the information from the data.} 
{A direct estimation of parameters is problematic: the likelihood function of the stochastic model is defined as a multiple integral over a high dimension event space.}
To the best of our knowledge, there exists no statistical method to estimate the parameters of such models.  As an approximation, we consider the likelihood of the {\color{black}deterministic differential system. We therefore estimate the parameters of the model  through its {\color{black}deterministic limit using experimental data of tumor growth in mice provided by \cite{Landsberg}}. 

The database is composed of tumor size measurements along time in three groups of mice: a control group showing the tumor growth in absence of treatment, a group treated with ACT therapy and a group treated with ACT therapy and restimulation, {\color{black}i.e.} reinjection of T cells.  {\color{black}We analyze these longitudinal data simultaneously with a Nonlinear Mixed Effects Model (NLMEM) \cite{pinero2000mixed} which takes into account the variability between mice (population and individual parameters are estimated), and estimate its parameters using a Stochastic Approximation Expectation Maximization (SAEM) algorithm \cite{convergenceEM_SAEM}.

Once the parameters have been estimated, T cell exhaustion is studied by heading back to the stochastic model. Indeed, complete exhaustion is a purely stochastic phenomenon which is not modeled by the deterministic system. As this can be a rare event, we estimate its probability using a splitting Monte Carlo method \cite{Morio, Damien}. {\color{black} {We observe that, in the range of parameters estimated from real data, the exhaustion probability can be non negligible, which confirms a conjecture in \cite{Loren}.} } 

{\color{black}As the relapse caused by the exhaustion of T cells can be delayed by their restimulation, our second objective is to optimize the ACT protocol: we compute the optimal treatment doses and restimulation times by minimizing  the T cell exhaustion probability at different stages of the disease evolution}.

{
The paper is organized as follows. 
In Section 2 we present experimental data which we use for the parameter estimation. 
In Section 3, the stochastic model and its deterministic limit are described. 
In Section 4, we first present the statistical tools involved in the parameter estimation of the deterministic system, the method to estimate the T cell exhaustion probability and finally a procedure to optimize the ACT treatment. 
Section 5 presents the results:  the estimated parameter values, the estimated  exhaustion probability  and the optimized treatment. The paper concludes with a discussion in Section 6.}

\section{Experimental data on Adoptive Cell Transfer therapy}
\noindent  
We first describe the experimental setup of \cite{Landsberg}, from which we use the data. Initially, {melanoma cells are injected in each of the 19 mice such that the typical initial tumor diameter is less than 1 milimeter ($\simeq 10^5$ cells). 
Mice are then split into 3 groups: untreated mice playing the role of a control group (denoted CTRL) and treated mice composed of two subgroups: mice treated once with ACT therapy (denoted ACT) and mice treated twice with ACT therapy (denoted ACT+Re).
After a time $t_a = 70$ days, 
mice of ACT and ACT+Re groups receive an intravenous delivery of $2\times 10^6$ active T cells which are able to kill differentiated melanoma cells. We suppose that $2\times 10^3$ stimulated T cells actually enter the tumor. The ACT+Re mice receive an additional injection of immunostimulatory nucleic acids in order to recover functions of exhausted T cells at time $t_{\text{Re}} = 160$ days. In our modeling, both restimulated and injected T cells are considered as active T cells. Thus, the restimulation reactivates $2\times10^3$ T cells.}

During the experiment, the tumor development is measured by palpation when the tumor is small and digital photography when the tumor is larger. The tumor size is measured weekly using a vernier caliper and recorded as mean diameter. {Longitudinal data for the 19 mice are composed of 5 mice of the CTRL group (with an average number of 10 observations {\color{black}per mouse}); 7 mice of the ACT group (with an average number of 26 observations); and 7 mice of the ACT+Re group  (with an average number of 33 observations).} 

The measure of tumor size, {\color{black}if it is smaller than} 2 mm, is very inaccurate due to the difficulties of palpation. Thus, these values are left-censored.
For tumor size between 2 and 3 mm, the inaccuracy is still important, about 1 mm. These values are interval-censored {\color{black}in} [2mm, 3mm]. 
Mice with tumor size exceeding 10 mm or showing signs of illness are killed. Thus measurements exceeding 10 mm are considered as right-censored.
Tumor sizes between 3 and 10 mm are considered to be measured correctly with a measurement accuracy {to be estimated}.

\label{database} 
Figure \ref{ShemaDonnees} represents the evolution of the tumor diameter along time for the 19 mice. The three censorships are illustrated by aligned horizontal points. Excluding the observations at $t_0 = 0$, mice in CTRL group are not concerned by left-censorship. Observe that the three groups of mice are distinguished by the speed of the tumor size evolution: tumor size of CTRL group (in blue) reaches 10 mm around the 100th day, tumor size of ACT group (in red) reaches 10 mm between 240 and 300 days and finally tumor size of ACT+Re group (in green) reaches 10 mm beyond the 300th day. Treatment and retreatment times are indicated by (black) vertical dotted lines. We  distinguish two phases in the ACT therapy (for treated mice): the growth phase (before $t_a = 70$) and the treatment phase (after $t_a = 70$). 
Note that the mice in CTRL group only experience the tumor growth phase. 

\begin{figure}
\begin{center}
  \includegraphics[height =0.38\linewidth, width=0.73\linewidth]{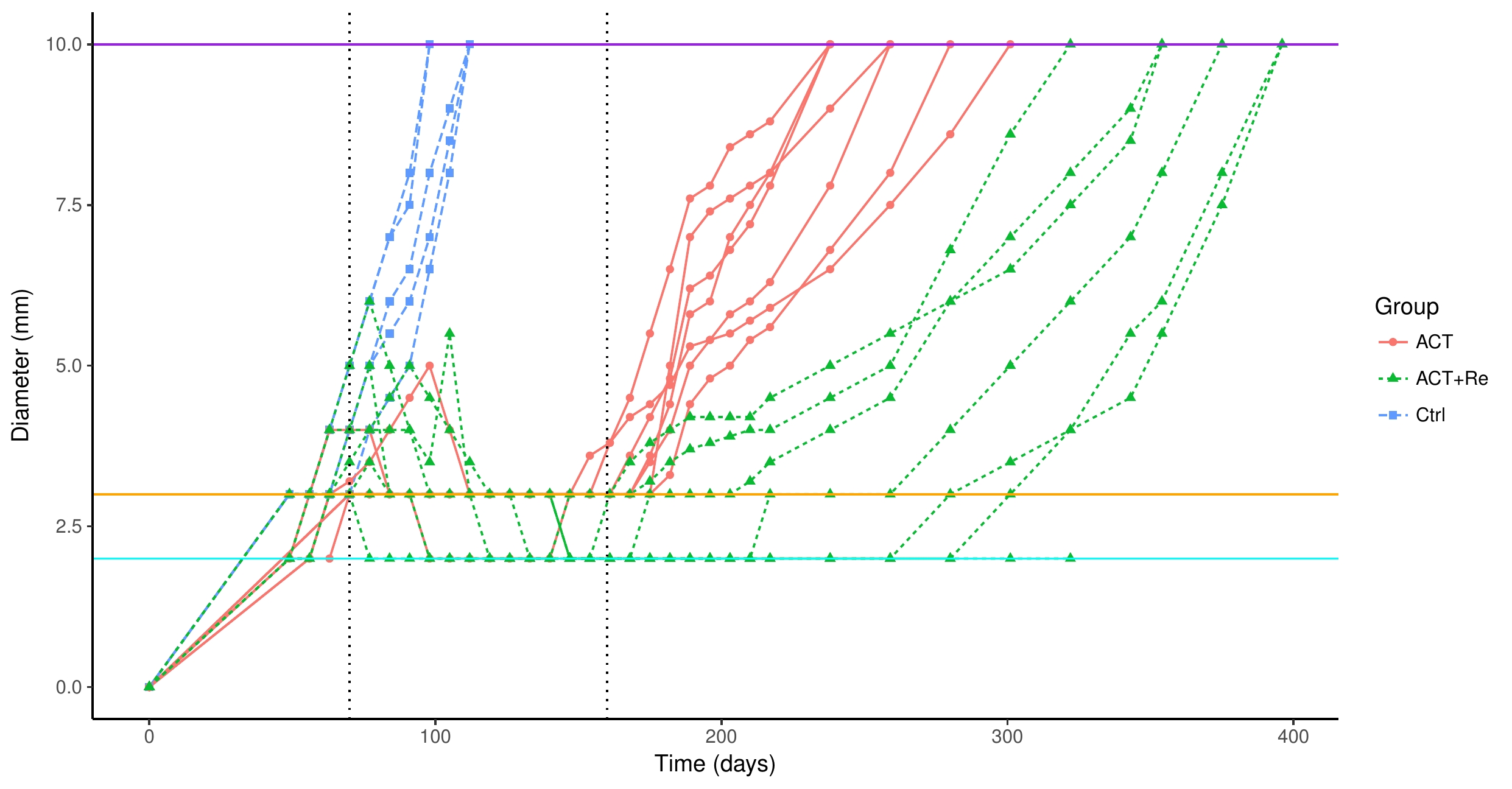}
  \caption{\scriptsize{Tumor diameter (mm) along time (days) in the groups CTRL (dashed line in blue), ACT (solid line in red), ACT+Re (dotted line in green). Censorship   indicated by   horizontal lines. Vertical dotted lines: Treatment (day 70) and   re-treatement (day 160) (for ACT+Re mice).}}
  \label{ShemaDonnees}
  \end{center}
\end{figure}

\section{Modeling tumor growth under treatment} 
\subsection{A stochastic model with four cell types} 
The stochastic model proposed in \cite{Loren} for the tumor growth under ACT therapy consists in a four-dimensional continuous time Markov process 
$$Z(t) = (M(t),D(t), T(t), A(t)) \in \mathds{N}^{4}$$ 
where population sizes at time $t$ are denoted by 
$M(t)$ for  differentiated melanoma cells, 
$D(t)$ for dedifferentiated melanoma cells, 
$T(t)$ for active T cells, and 
$A(t)$ for cytokines $TNF_{\alpha}$. 
{The carrying capacity $K$ of the system is assumed to be equal to $10^5$ cells, which corresponds to a tumor diameter of the order of the milimeter. This parameter scales the interaction rates between different cell populations, since the interaction is assumed to be mean-field (each cell interacts with all others, no space dependency is considered.)}

{The main features of this stochastic individual-based process are the microscopic modeling of 
the {proliferation} 
of cancer cells and active T cells, 
the secretion of cytokines TNF-$\alpha $, and 
the death, exhaustion, and clearance of, respectively, cancer cells, active T cells, and cytokines TNF-$\alpha$.
Those events are modeled with non-linear birth and death processes including a predator-prey interaction between cancer cells and T cells. 

The complex biological process of T cell exhaustion is chosen to be modeled (and summarised) as the death of active T cells. More precisely, the time between the activation of a T cell and its exhaustion is assumed to be an exponentially distributed random variable.

Another mechanism which is not explicitly modeled concerns interferons IFN-$\gamma$ which are a by-product of inflammation and strongly upregulate the expression of the markers on melanoma cells (see Supplementary Fig. 15 in \cite{Landsberg}). Biologically, IFN-$\gamma$-dependent markers upregulation and TNF-$\alpha$-dependent dedifferentiation represent two functionally connected inflammation-induced adaptive mechanisms that together contribute to acquire resistance to ACT therapy \cite{Landsberg}.  
In the model of Baar et al.\ \cite{Loren}, the role of interferons IFN-$\gamma$ is qualitatively taken into account in the killing rate of differentiated melanoma cells by T cells.
This significant simplification of real mechanisms underlying T cell exhaustion allows to work with a mathematically tractable model. 
}

In the growth phase of the tumor ($t < 70{\text{ days}}$) only division, switch and death of melanoma cancer cells happen. At this stage, there are no active T cells nor cytokines $TNF_{\alpha}$ : $(T(t),A(t))=(0,0)$. The evolution of $M(t)$ and $D(t)$ is modeled by birth and death processes \cite{kendall1948generalized} including additional terms modeling the switches between the two cancer cell populations.  
Initial conditions are $(M(0),D(0))=(M_0,0)$.  

At time $t = 70$ days (beginning of treatment), {a dose of $2000$} active T cells is injected in the system: $T(70) = 2000$. Then, {\color{black}all cell populations and cytokines} ($M(t)$, $D(t)$, $T(t)$, $A(t))$ evolve starting with initial conditions ($M(70)$, $D(70)$, $ 2000 $, $ 0)$. 
 The system models the evolution of differentiated melanoma cells $M(t)$ (preys) in the presence of active T cells $T(t)$ (predators) as a predator-prey framework. An additional natural exhaustion of active T cells is modeled by a simple death process. A deterministic {secretion} of cytokines $A(t)$ occurs at each T cell division event while the clearance of cytokines is modeled by a death process. An additional switch from differentiated to dedifferentiated melanoma cell due to the presence of cytokines is also modeled. The ACT group is modeled by the growth phase and this first treatment phase. 

At time $t = 160$ days, an additional dose of T-cell stimulant is injected for mice in ACT+Re group re-activating 2000 T cells. 
New initial conditions are ($M(160)$, $D(160)$, $T(160) + 2000$, $A(160))$. 

Figure  \ref{Dynamics} summarizes the dynamics of the process during the treatment phase (see Section 2.2 of  \cite{Loren} for more details). 
%
\begin{figure}
\begin{center}
  \includegraphics[width=0.5\linewidth, height=0.33\linewidth]{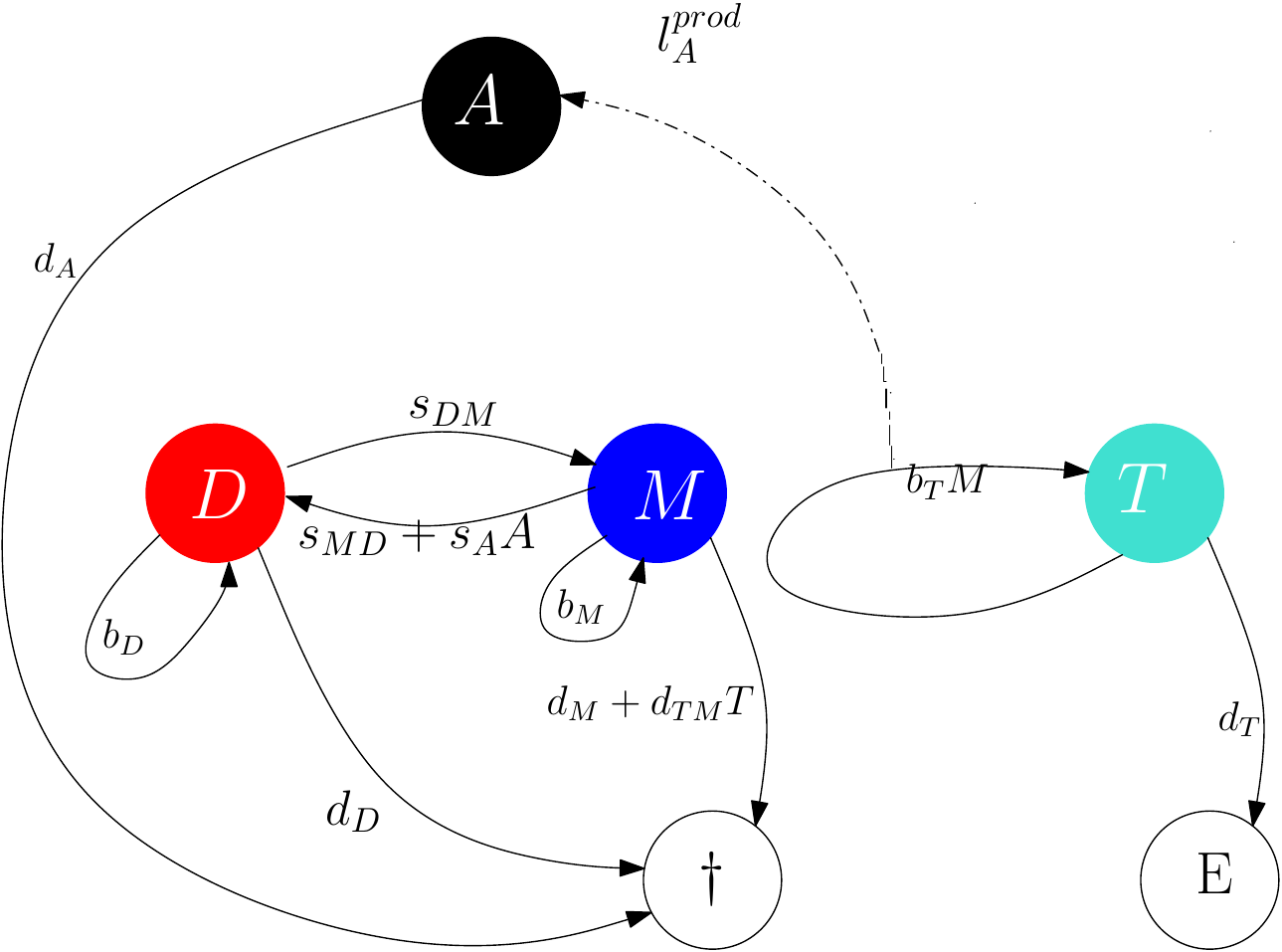}
  \caption{\scriptsize{{Melanoma stochastic process: 
  $M$: differentiated melanoma cells;  $D$: dedifferentiated melanoma cells; $T$: T cells; $A$: cytokines $TNF_\alpha$. $\dagger$: death of cancer cells and   clearance of cytokines; $E$: T cells exhaustion. 
 Arrow with same initial and end points  indicates a division   in this population. Values above the arrows represent the rate parameters.}}}
  \label{Dynamics}
  \end{center}
\end{figure}
All   events in the stochastic model (division, death and switch) occur after {random} exponential waiting times regulated by rate parameters. Parameters $b_M$ and $b_D$ (Figure \ref{Dynamics}) thus represent division rates of differentiated (respectively dedifferentiated) melanoma cells and correspond to the inverse of the average waiting time before observing the division of a differentiated (respectively a dedifferentiated) melanoma cell. Similarly, $d_M$ and $d_D$ correspond to natural death rates of differentiated (respectively dedifferentiated) melanoma cells, while $s_{MD}$ and $s_{DM}$ represent respectively the switching rates from differentiated melanoma cells to dedifferentiated one and the {\color{black}converse}. Parameters $b_T$ and $d_T$ are respectively the division and exhaustion rates of active T cells. To a certain extent, $b_T$ can be seen as an effective parameter corresponding to $b_T = b_{T}^0 - d_{T}^0$ with $b_{T}^0$ and $d_{T}^0$ respectively the therapy induced division and exhaustion rates of active T cells. Indeed, differentiated melanoma cells secrete a substance that exhausts active T cells with a rate $d_{T}^0$ in addition to their natural exhaustion rate $d_T$. {The therapy induced death of differentiated melanoma cells} is regulated by the rate parameter $d_{MT}$ while   parameter $d_A$ corresponds to the clearance rate of cytokines $TNF_{\alpha}$. {\color{black}The} product $l_{A}^{\text{prod}} \times b_T$ represents {the secretion rate of cytokines induced by} {\color{black}the} therapy, and $s_A$ is associated to the additional switching from differentiated to dedifferentiated melanoma cells induced by cytokines. 
Parameters  are recapitulated in Table \ref{table:tableParamModeleImmuno} of the Appendix. 


\subsection{Deterministic limit of the stochastic model}
The rescaled stochastic model converges in the limit of large population of cells ($K \to \infty$) to the solution of the following deterministic differential system:   
\begin{equation}
\left\{
\begin{array}{l}
  \mathbf{\dot{n}_{M}} = (b_M - d_M)  \mathbf{n_{M}}  - s_{MD}  \mathbf{n_{M}}  + s_{DM}  \mathbf{n_{D}}- s_A  \mathbf{n_A n_{M}} - d_{MT} \mathbf{n_T}  \mathbf{n_{M}}  \\
   \mathbf{\dot{n}_{D}}   \hspace{0.1em} = (b_D  \hspace{0.2em} -   \hspace{0.3em} d_D)  \mathbf{n_{D}} + s_{MD}  \mathbf{n_{M}} - s_{DM}  \mathbf{n_{D}} + s_A  \mathbf{n_A n_M}  \\
   \mathbf{\dot{n}_{T}} = - d_T  \mathbf{n_{T}}   + b_T  \mathbf{n_M n_{T}} \\
   \mathbf{\dot{n}_{A}} = - d_A \mathbf{ n_A}  + l_{A}^{\text{prod}} b_T  \mathbf{n_M n_T}
\end{array}
\right.
\label{SystemModelComplet}
\end{equation}

\noindent with initial condition $(n_{M_0}, n_{D_0}, n_{T_{0}}, n_{A_{0}})$. 
Quantity $\mathbf{n_X}(t) \in \mathds{R}$ (with $X = M, \ D, \ T \text{ or }  A$) represents the quantity of $X$ at time $t$ and $\mathbf{\dot{n}_X}$ represents the variation of this quantity along time.
Note that we work in this paper on a simplified version of the initial model \cite{Loren}. Some competition terms were removed as they cannot be estimated from the data. Indeed, mice are killed before the tumor size (number of cancer cells) reaches the (non trivial) fixed point regulated by the competition terms.  To avoid the problem of identifiability, we set $d_M = d_D = 0$ since only the difference between division and death rates of melanoma cells can be estimated from the data. 

{
The different phases of the therapy are similar to the ones of the stochastic model.} During the growth phase ($t < 70$), only populations of cancer cells $n_ {M}(t)$ and $n_{D}(t)$ evolve $(n_{T}(t),n_{A}(t))=(0,  0)$. All parameters associated with the treatment are then equal to zero. At time $t = 70$, the quantity of active T cells is set as $n_{T}(70) = d_{70}$. The four populations of cells ($ n_ {M}(t) $, $ n_{D}(t) $, $n_{T}(t) $, $ n_{A}(t))$ evolve. Model \eqref{SystemModelComplet} corresponds to the dynamics of the ACT group.  For ACT+Re group, an additional quantity of active T cells is added in the system at time $t_{\text{Re}} = 160$ ($n_{T}(160) = n_ {T}(t) + d_{\text{Re}}$).

\section{Methods}
\subsection{Mixed Effects Model for tumor growth under treatment}
Parameters are estimated by analyzing the three groups simultaneously. We use a {\color{black}Nonlinear} Mixed Effects regression Model (NLMEM) to take into account the inter-mice variability \cite{pinero2000mixed}. Parameter estimation with NLMEMs is difficult since the likelihood function does not have an explicit form. When the regression function of   NLMEM is a stochastic model, the high dimension of the event space prevents to compute the likelihood. Thus, {\color{black}we choose to perform the parameter estimation based on}   the deterministic model (\ref{SystemModelComplet}).

Let us define $y_i = (y_{i1},\dots,y_{in_i})$ where $y_{ij}$ is the (noisy) measurement of the tumor diameter   for mouse $i$ at time $t_{ij}$,
$i = 1,\dots,N$, $j = 1,\dots,n_i$, and set $y = (y_1,\dots,y_N)$. The NLMEM is defined by:
\begin{eqnarray}
	y_{ij} = f(\psi_i,t_{ij}) +\epsilon_{ij},\\
 \psi_i = \mu {\color{black} \odot} \tau_i \exp \{ {\color{black}\eta_i \odot \tau_i}\} ,  \nonumber
\label{NLMEM_cancer}
\end{eqnarray} 

\noindent
where $\odot$ denotes component-wise multiplication and
{\color{black}
\begin{equation*}
\begin{array}{l}
\vspace{0.1cm}
\mu \hspace{0.1cm}= (b_{M},b_{D}, s_{MD}, s_{DM}, n_{M_0}, b_T, d_T, d_{MT}, s_A, d_A, l_{A}^{\text{prod}}) \ \text{  is a vector of fixed effects; } \\
\vspace{0.1cm}
\tau_i = (\hspace{0.1cm}1, \hspace{0.25cm} 1, \hspace{0.5cm}1, \hspace{0.5cm}1,\hspace{0.5cm}1, \hspace{0.4cm}0, \hspace{0.3cm}0, \hspace{0.2cm} 0, \hspace{0.2cm}0, \hspace{0.3cm}0, \hspace{0.2cm}0 \hspace{0.4cm}) \text{ when $i$ is a control individual; } \\
\tau_i = (\hspace{0.1cm}1,\hspace{0.25cm}1,\hspace{0.5cm}1,\hspace{0.5cm}1,\hspace{0.5cm}1,\hspace{0.4cm}1,\hspace{0.3cm}1,\hspace{0.2cm}1,\hspace{0.2cm}1,\hspace{0.3cm}1,\hspace{0.2cm}1\hspace{0.4cm}) \text{ when $i$ is a treated individual; }\\
 \epsilon_i =(\epsilon_{i1},\dots,\epsilon_{in_i}) \sim \mathcal{N}(0, \sigma^2 I_{n_i})  \text{ represents the residual error}\\
\eta_i = ({\color{black}\eta_{b_{M}}^{i}}, {\color{black}\eta_{b_{D}}^{i}}, {\color{black}\eta_{s_{MD}}^{i}}, {\color{black}\eta_{s_{DM}}^{i}}, {\color{black}\eta_{n_{M_{0}}}^{i}},  {\color{black}\eta_{b_T}^{i}}, {\color{black}\eta_{d_T}^{i}}, {\eta_{d_{MT}}^{i}}, {\color{black}\eta_{s_A}^{i}}, {\color{black}\eta_{d_A}^{i}}, {\color{black}\eta_{l_A}^{i}}) \sim \mathcal{N}(0,\Omega)
\end{array}
\label{where_NLMEM}
\end{equation*}}

\noindent $\eta_i$ is a vector of random effects independent of $\epsilon_i$; $\psi_i$ represents the vector of individual regression parameters; 
$$ f(\psi_i,t) = (n_{M}(\psi_i,t) + n_{D}(\psi_i,t))^{\frac{1}{3}}$$
  describes the tumor diameter (up to some constant), with $n_{M}$ and $n_{D}$ being solution of the nonlinear system (\ref{SystemModelComplet});  $\sigma^2$ is the residual variance; $I_{n_i}$ the identity matrix of size $n_i$; $\Omega$ the variance matrix of the random effects quantifying variability between mice.
We note $\theta = (\mu, \Omega, \sigma^2)$ the parameters to be estimated. 

Note that a biological constraint states that the natural switch rate from a dedifferentiated melanoma cell to a differentiated melanoma cell ($s_{DM}$) is larger than {\color{black}the rate in the other direction} ($s_{MD}$) \cite{Landsberg}. This constraint is introduced in the estimation procedure.

\subsection{Parameter estimation using SAEM algorithm}

The SAEM-MCMC algorithm which combines the Stochastic Approximation Expectation Maximization algorithm \cite{convergenceEM_SAEM} with a Markov chain Monte Carlo procedure adapted to censored data \cite{AdelineHdr, Kuhn}, implemented under the software MONOLIX \cite{Monolix}, is used to estimate the model parameters. 
We start with the model having random effects on all parameters and then use likelihood ratio tests (LRT) to select the significant random effects. Then,  the effect of the categorical covariate group  denoted by $\mathcal{G}$ is tested on the fixed effects $\mu$ through LRT. The  ACT group is taken as reference group, $\mathcal{G}$ = $\{$ACT*, ACT+Re, CTRL$\}$. Standard errors of the estimated parameters of the selected models are estimated through the computation of the Fisher Information Matrix. %

\subsection{T cell exhaustion probability estimation using Monte-Carlo and Importance Splitting algorithms}

T cell exhaustion is a cause of relapse   \cite{Loren}. It is defined by the {\color{black} stochastic} event $\{T(t) \leq S, \ t \le t_F\}$ where $S$ is the exhaustion threshold and $t \le t_F$ a condition for the exhaustion to occur in a finite time. The {\color{black}complete} T cell exhaustion ($S = 0$) is a phenomenon which can only occur in the stochastic system: when the T cell population is low enough, the stochastic fluctuations can drive it to extinction, whereas the T cell population can never vanish in the  deterministic limiting system. {\color{black}This follows from the analyticity of the solutions to \eqref{SystemModelComplet}, given that $T(70)>0$.} 

We want to estimate the exhaustion probability  $p = \mathds{P} (T(t) \le S, \ t \le t_F)$. 
Let $u  =  \mathds{1}_{T(t) \le S}$.  
Using {\color{black}the intuitive} Monte Carlo method, the T cell exhaustion probability is estimated by $\hat{p} = {\sum_{k = 1}^{N_T} u_k}/{N_T}$ with $N_T$ the total number of simulations. {\color{black}However, depending on the values of the model parameters, the exhaustion threshold $S$ may be difficult to reach leading   to very small probabilities. Monte Carlo method thus requires a large   $N_T$  when the probability $p$ is very small. This method is therefore not suitable for estimating the probability of T cell exhaustion because the variance will diverge when the probability tends to zero for a reasonable value of $N_T$  \cite{Morio}.} To reduce the variance for very small probabilities estimation, an alternative is the Importance Splitting (IS) algorithm designed for rare event probability estimation \cite{Morio, Damien, cerou2007adaptive}. 
{\color{black} Indeed}, IS gradually calculates the probability of reaching the threshold $S$ (the rare event) through the calculation of the probability of reaching intermediate thresholds easier to reach than $S$. {\color{black}Thus,} the probability $\mathbb{P}(T(t) \le S, \ t \le t_F)$ is calculated according to the splitting principle by 
\begin{equation}
\mathbb{P}(T(t) \le S, \ t \le t_F) = \prod_{k = 1}^{m} p_k
\end{equation}
\noindent with $p_1 = \mathbb{P}(T(t) \le S_1, t \le t_F)$  and $p_k = \mathbb{P}(T(t) \le S_k \mid T(s) \le S_{k-1}, \ s \le t \le t_F)$ for $k = 2,\dots,m$. The $S_k$ ($k < m$) are the intermediate thresholds and $S_m = S$ the exhaustion threshold. 

At each iteration $k$, the IS algorithm consists in   simulating $N_{T}$ trajectories of   process $T(t)$ and   estimating    the intermediate probability $p_k$ by considering the   $N_{S_k}$   trajectories which reach the intermediate threshold $S_k$ before time $t_F$: $\hat p_k = \frac{N_{S_k}}{N_{T}} $. 
To pass from iteration $k$ to iteration $k+1$, $N_{T}$ trajectories are sampled from the $N_{S_k}$ trajectories having reached the threshold $S_k$ (by allowing replacement in the sampling) and run the $N_{T}$ new trajectories in order to reach the next threshold $S_{k+1}$. 
In our setting, T cells are completely exhausted at time $t$ if the quantity $T(t)$ has reached $S_5 = 0$. 
In   Importance Splitting algorithms, it is possible to place the intermediate 	thresholds adaptively using a quantile method. We can also place these 	thresholds "manually" when we have enough knowledge about the dynamics of the 	stochastic system, which is the case for the dynamics of T cells. 
	In this paper, we tested different numbers (3, 4 and 5) and spacing of intermediate thresholds 	until the  numerical value of the exhaustion probability become stable from one series of  1000 simulations to another. We then fix 	these thresholds to $S_1 = 7 \times 10^{-5}$ (or $0.35\%$ of the initial quantity of active T cells), $S_2 = 5 \times 10^{-5}$ ($0.25\%$ of $T(70)$), $S_3 = 3 \times 10^{-5}$ ($0.15\%$ of $T(70)$), and $S_4 = 2 \times10^{-5}$ ($0.1\%$ of $T(70)$). We set $N_T$ to $1000$.  

The computation of the T cell exhaustion probability   requires {\color{black}$N_T = 1000$ simulations of the stochastic model for each of the $m = 5$ thresholds. Furthermore, each trajectory can contain up to $100$ millions of stochastic events. This stochastic approach is therefore very expensive in computing time. For example, the average time to calculate the exhaustion probability   is approximately $100$ hours {\color{black}(using $3$ cores, with 7GB of memory per core)}}.

However, this stochastic approach allows us to highlight a link between the probability of T cell exhaustion and the depth of the first minimum of the T cell deterministic trajectory. Indeed, since the stochastic process is locally close to the deterministic system plus a Gaussian noise \cite{Ethier}, the lower the deterministic minimum the larger the probability of T cell exhaustion. 
We exploit this relation to ease the study of the impact of remaining parameters which have a random effect on the T cell exhaustion phenomenon. 

\subsection{Criteria to optimize treatment doses and restimulation times}\label{sec:OptValComput}

The kind of relapse we can act on without changing the medical setup is the relapse due to T cell exhaustion. The goal is thus to minimize the probability of T cell exhaustion, {which amounts in the deterministic setting to maximize the value of the first minimum of the T cell population}.
Treatment parameters to be optimized are the treatment dose $d_{70}$, the retreatment time $t_{\text{Re}}$ and the retreatment dose $d_{\text{Re}}$. 
The cost function (to be maximized) is thus defined by 
\begin{equation}
\begin{array}{l}

g = \minimum \{n_T(t), t \le t_F \}

\end{array}
\label{g_min_det}
\end{equation}
\noindent with $n_T(t) $ the solution of deterministic system (\ref{SystemModelComplet}) and $t_F$ a fixed finite time. Note that $g$ depends on all the model parameters through $n_T(t)$. Its unit is a ${\text{number of T cells}}/{K}$.}

 We optimize the therapy by maximizing $g$ over different subsets of treatment parameters. 
{We refer to $g$ as the minimum of the T cell trajectory, or simply, as the T cell minimum.}

{To take random effects into account in the optimization, we consider   specific quantiles of their distribution. We focus on the T-cell related parameters $d_T$ (natural death rate of T cells), $d_{MT}$ (therapy induced death rate of the differentiated melanoma cells) and $l_{A}^{\text{prod}}$ (which, multiplied by $b_T$, leads to the therapy induced {secretion} rate of the cytokines). 
We then set the value of the natural clearance rate $d_A$ of cytokines to its population value since it is less related to T cells compared to the three previous treatment parameters, the cost function $g$ is almost constant as a function of $d_A$. 
We also set the value of the disease parameters $b_M$ and $b_D$ to their population mean assuming that these are disease-specific parameters that are less under control. 
{As $n_{M_0}$ corresponds to the initial size of the tumor,  we   consider different quantiles of its distribution.} 
In summary, we set $ b_M $, $ b_D $ and $ d_A $ to their population values and consider  quantiles of order 5\%, 50\%, 95\% for $d_T$, $d_{MT}$, $l_{A}^{\text {prod}}$ and $n_{M_0}$. }
{These sets of quantiles for the variables $\psi_i$ are denoted   $\mathcal Q(\psi_i)$ and are provided in Table \ref{table:tableResu}.}
{Our choice of quantiles give a better idea on the dispersion of the estimated parameters in their estimation range compared, for example, to the quartiles}. 
%
{In total, we consider $3^4 = 81$ sets of parameter values corresponding to {three quantiles} for each of the four parameters $d_{MT}, \ d_T, \ l_{A}^{\text{prod}}$, and $n_{M_0}$}.

We consider two optimization procedures: 

\begin{itemize}
    \item the first is performed on the ACT group and we focus on the dependence of the function $g$ on the treatment dose $d_{70}$. The optimal dose $d_{70}^{\text{opt}_{*}}$ is defined as follows:
    
\begin{equation}
d_{70}^{\text{opt}_{*}} = \argmaxA_{d_{70}  \in [d_{\text{min}}, d_{\text{max}}]}  g(d_{70}).
\label{optimizationACT}
\end{equation}

    \item the second optimization procedure is performed on the ACT+Re group and we focus on the dependence of the function $g$ on the treatment dose $d_{70}$, the restimulation dose $d_{\text{Re}}$ and the restimulation time $t_{\text{Re}}$. The optimal treatment dose {\color{black}$d_{70}^{\text{opt}_{**}}$}, retreatment dose $d_{\text{Re}}^{\text{opt}}$ and retreatment time $t_{\text{Re}}^{\text{opt}}$ are defined as follows:
\begin{equation}
[d_{70}^{\text{opt}_{**}}, d_{\text{Re}}^{\text{opt}}, t_{\text{Re}}^{\text{opt}} ] = \argmaxA_{(d_{70}, d_{\text{Re}}, t_{\text{Re}}) \in \mathcal{D}}  g(d_{70}, d_{\text{Re}},t_{\text{Re}}), 
\label{optimizationACTRE}
\end{equation}

\noindent {\color{black}where $\mathcal{D}$ = $[d_{\text{min}}, d_{\text{max}}] \times [d_{\text{min}}^{\text{Re}}, d_{\text{max}}^{\text{Re}}] \times [t_{\text{min}}^{\text{Re}}, t_{\text{max}}^{\text{Re}}]$.} 

\end{itemize}

We set the following initial values:
$t_0 = 0$; 
$n_{M_0}  \in \mathcal Q(n_{M_{0_i}})$ indicating different initial quantities of differentiated melanoma cells; $d_{70}^{\text{initial}} = 0.02$ the initial stimulation dose, $d_{\text{Re}}^{\text{initial}} = 0.02$ the initial restimulation dose and $t_{\text{Re}}^{\text{initial}} = 160$ the initial restimulation time.

The treatment dose optimization interval $[d_{\text{min}}, d_{\text{max}}]$ is set to $[0.005, 0.04]$ for both ACT and ACT+Re groups and the retreatment dose optimization interval $[d_{\text{min}}^{\text{Re}}, d_{\text{max}}^{\text{Re}}]$ is set to $[0, 0.04]$ for ACT+Re group {\color{black}(such that $d_{\text{Re}}^{\text{opt}} = 0$ when $d_{70}^{\text{opt}_{**}}$ is sufficient to avoid T cell exhaustion before time $t \le t_F$)}.  {\color{black}Indeed, from an experimental point of view, the initial dose $d^{\text{initial}} = 0.02$ was chosen to have a good control on the tumor} \cite{Meri}. 
This leads us to set $d_{\text{min}}$, $d_{\text{max}}$, $d_{\text{min}}^{\text{Re}}$ and $d_{\text{max}}^{\text{Re}}$ not too far from their initial values. Similar considerations lead us to set the retreatment time optimization interval to $[t_{\text{min}}^{\text{Re}} , t_{\text{max}}^{\text{Re}}] = [130, 190 ]$. Time $t_{F}$ is set to $t_{F} = 300$ in the ACT group and to $t_{F} = 400$ in the ACT+Re group {\color{black}(based on experimental data of Figure \ref{ShemaDonnees})}.


The optimal treatment parameter values (doses and restimulation time) are computed from the plots of the function $g$ over intervals of discrete values of doses and restimulation time, for fixed sets of $\{d_{MT}, \ d_T, \ l_{A}^{\text{prod}}, \ n_{M_0}\}$ taken in the estimation range. 

\section{Results}
\subsection{ No effect of the covariate group on the parameters}
Likelihood ratio tests  lead to the model given by $\{ {\color{black}\eta_{r_{M}}^{i}} \ne 0, {\color{black}\eta_{r_{D}}^{i}} \ne 0, {\color{black}\eta_{l_A}^{i}} \ne 0, {\color{black}\eta_{d_{MT}}^{i}} \ne 0,{\color{black}\eta_{d_T}^{i}} \ne 0, {\color{black}\eta_{d_A}^{i}} \ne 0, {\color{black}\eta_{n_{M_{0}}}^{i}} \ne 0\}$. 
LRT also showed that none of the parameters is function of the covariate group $\mathcal{G}$,  i.e., the model captures the differences between groups well enough without the help of an additional covariate.
Thus, we take as final model the one without covariates. Table \ref{table:tableResu} contains the estimates and the associated standard errors of the final model, Figure \ref{fits_indiv_final} {\color{black}(and Figure \ref{fits_all_inds_final} in Appendix)} 
shows its individual fits. 
We believe the standard errors, as well as the other validation elements in the Appendix like Figure \ref{pred_vs_obs_final} (observations versus predictions),  or Figure \ref{pcVPC_final} (Visual Predictions Check (VPC)) to be satisfactory in view of the small number of data  used for the estimation.
Confidence in the model predictions might be improved if model predictions could be related to experimentally measured parameters, like the expected ratios of T cells over tumor cells, or the ratios of antigen expressing over antigen loss tumor cells.  
However, we do not have access to such experimental data.

{\small
\begin{table} 
\centering
\begin{tabular}{llll ll}
  \hline
   & $\mu$ estimation  &    $\omega$ estimation & Quantile & Quantile\\
  &   (r.s.e (\%)) &      (r.s.e (\%)) & 5\% & 95 \% \\
  \hline
   \vspace{0.cm}
   $b_{M}$  & 0.09 (4) &  0.10 (34)   \\
    \vspace{0.1cm}
   $b_{D}$ & 0.05 (10)  &  0.35 (20)  \\
   
    \vspace{0.1cm}
   $s_{MD}$ & $< 0.01$ (118)  &  0 (-)   \\
    \vspace{0.1cm}
    
$s_{DM}-s_{MD} $& $< 0.01$ (-)  & 0 (-)   \\
   
  \vspace{0.2cm}
  $n_{M_0}$ & 0.08 (28)  &  0.58 (29)  & 0.03   &  0.21 \\
  
    \vspace{0.cm}
   $b_T$ & $<0.01$ (16)  &  0 (-)   \\
     \vspace{0.1cm}
   $d_{T}$ & 0.02 (34)  &  1.07 (27) & $< 0.01$    & 0.09 \\
    \vspace{0.1cm}
   $d_{MT}$ & 1.33 (55)  &  1.03 (48)   & 0.24   &  7.24 \\

 \vspace{0.1cm}
   $s_{A}$ & 77 (37)  &  0  (-)   \\

 \vspace{0.cm}
   $d_{A}$ & 0.03 (-)  &  0.09 (181)  \\

    \vspace{0.2cm}
   $l_{A}^{\text{prod}}$ & 0.19 (93)  &   3.23 (25) & $<0.01$     &  39.37  \\ 
      
  $\sigma$ & 0.44 (6)  &     \\
  \hline
  \end{tabular}
 \caption{\scriptsize{Estimated parameters for the \textbf{final model}: population mean $\mu $ and inter-mice standard deviation $\omega$. r.s.e $= \frac{\text{standard error}}{\text{estimation}}  \times 100$ (in brackets). $\omega = 0$ for parameters without random effect, and (-)  when the standard error is not estimated.  Quantile values for  random effects considered in treatment optimization are given in the last two columns. 
 } 
 }  
  \label{table:tableResu}
  \end{table}
}

%

\begin{figure}
     \centering
     \subcaptionbox{CTRL group\label{1}}{\includegraphics[width = 0.45\textwidth]{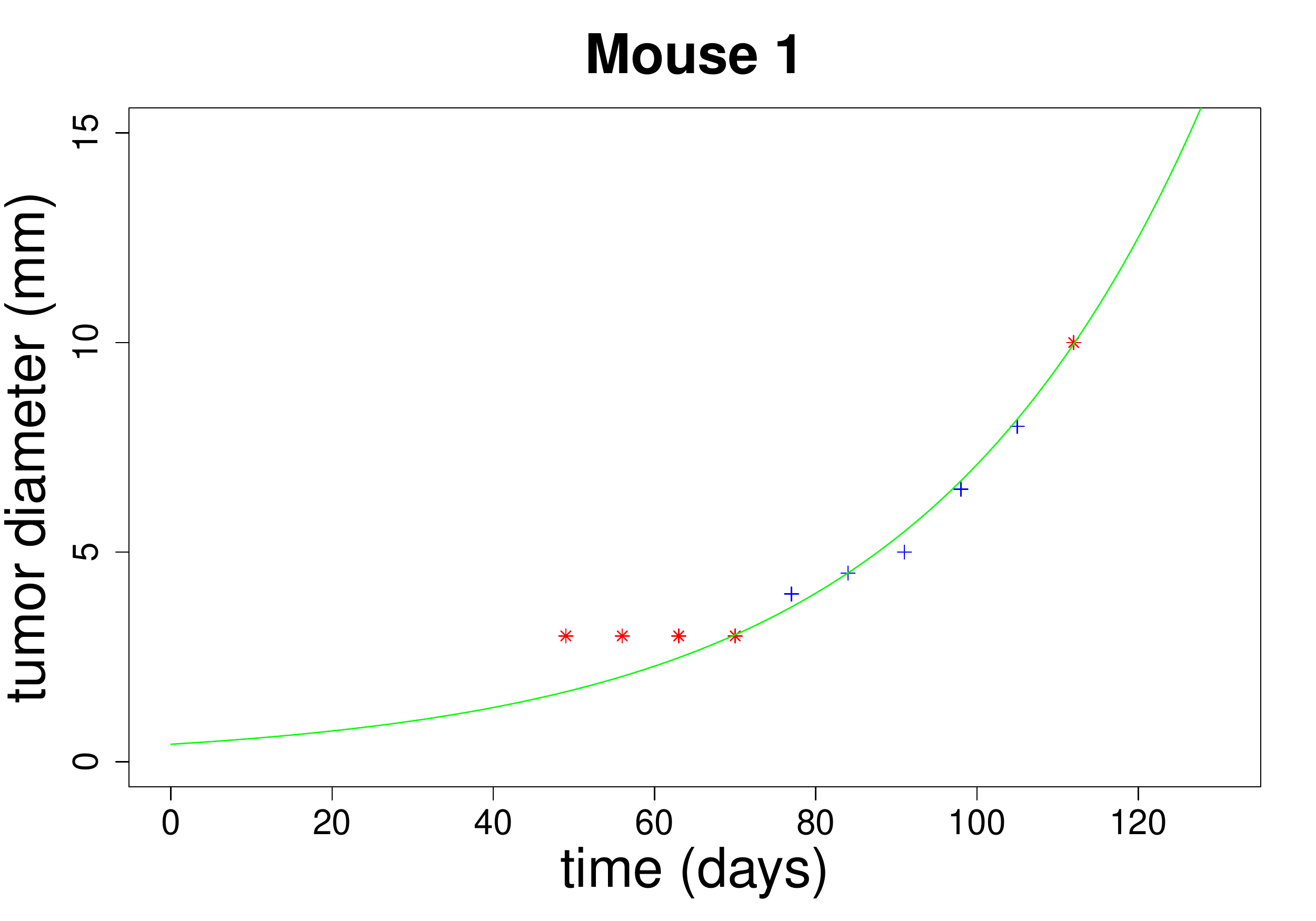}}\quad
     \subcaptionbox{ACT group\label{2}}{\includegraphics[width = 0.45\textwidth]{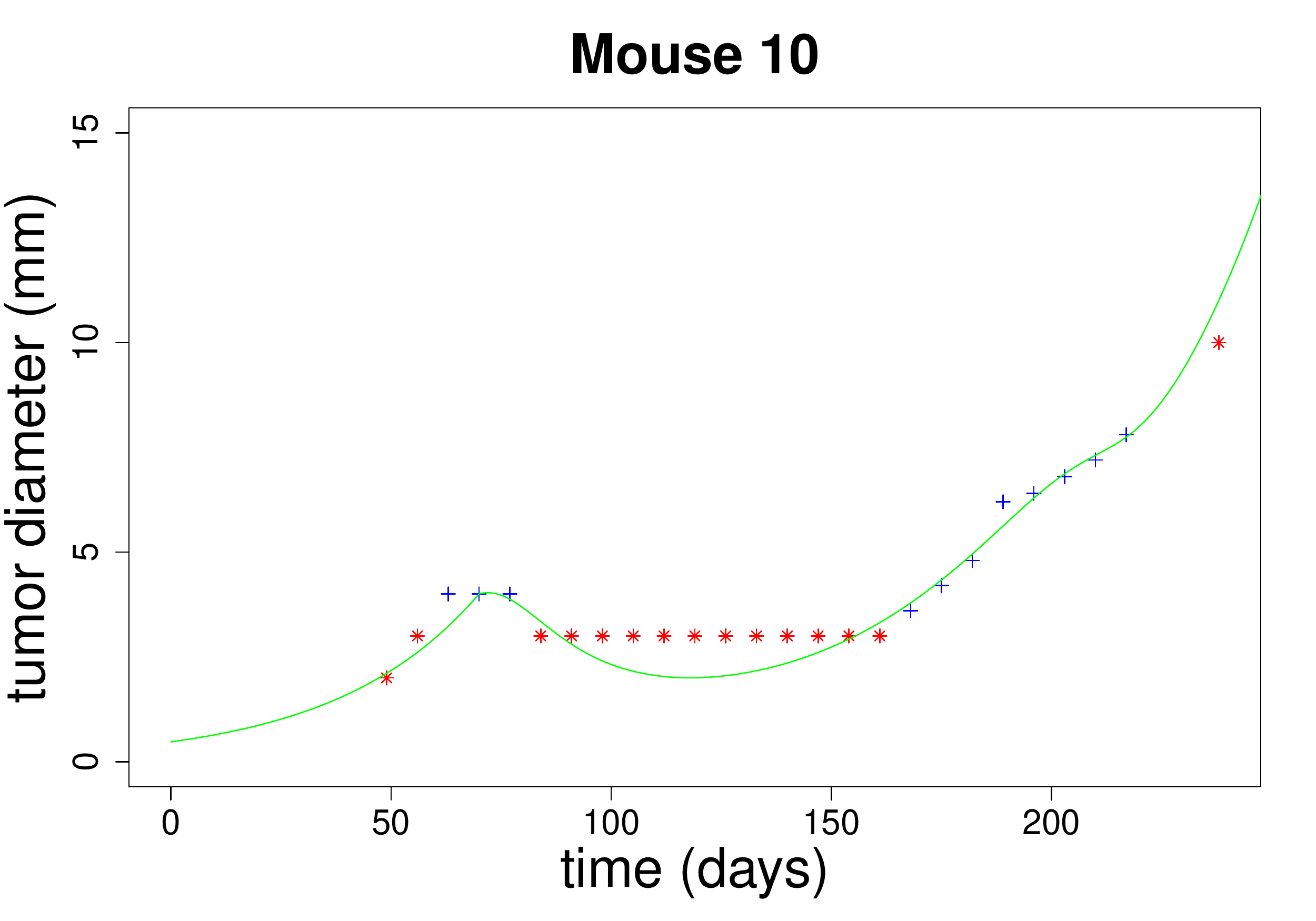}}\quad
     \subcaptionbox{ACT+Re group\label{3bis}}{\includegraphics[width = 0.45\textwidth]{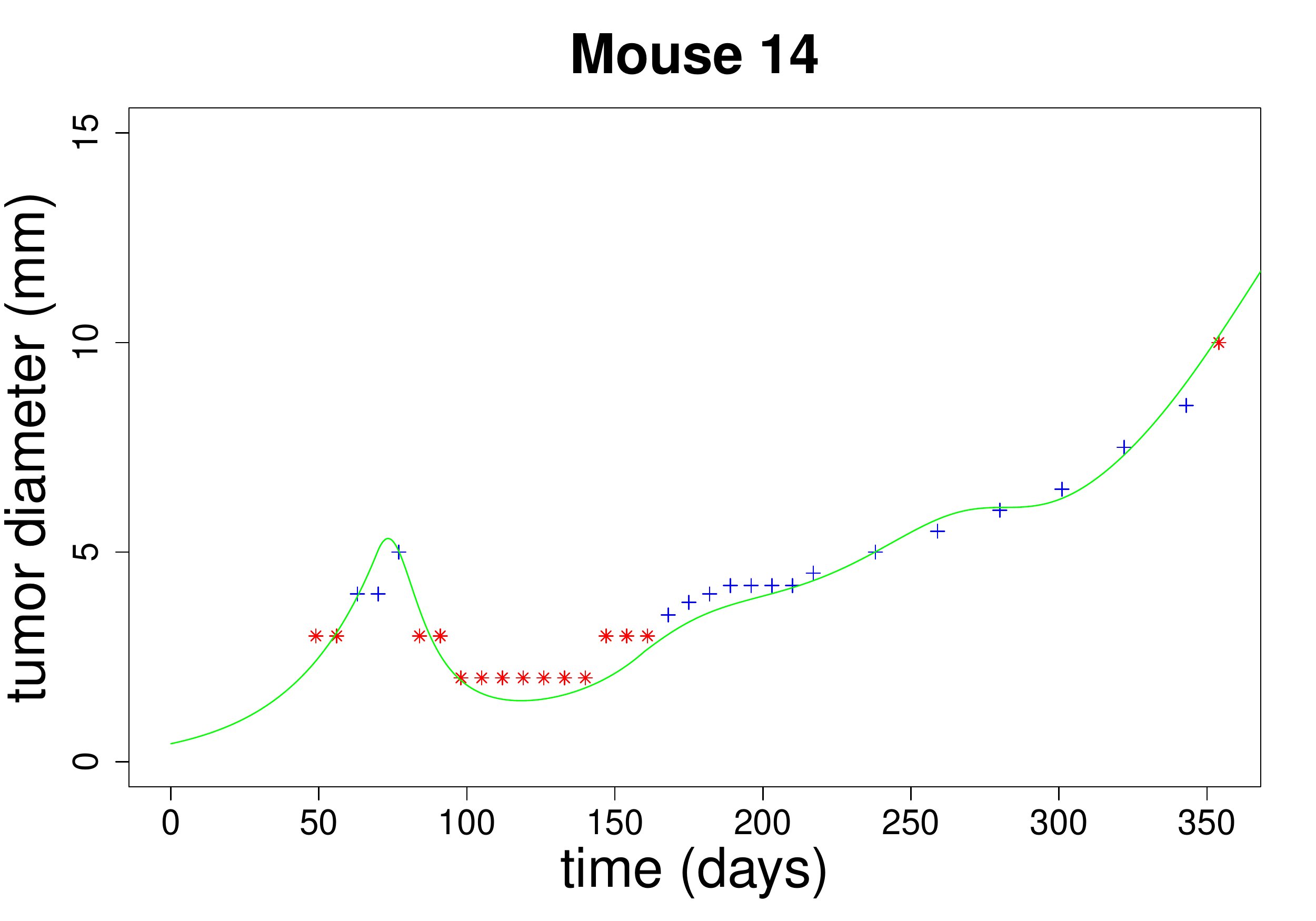}}\quad
     \subcaptionbox{ACT+Re group\label{3}}{\includegraphics[width = 0.45\textwidth]{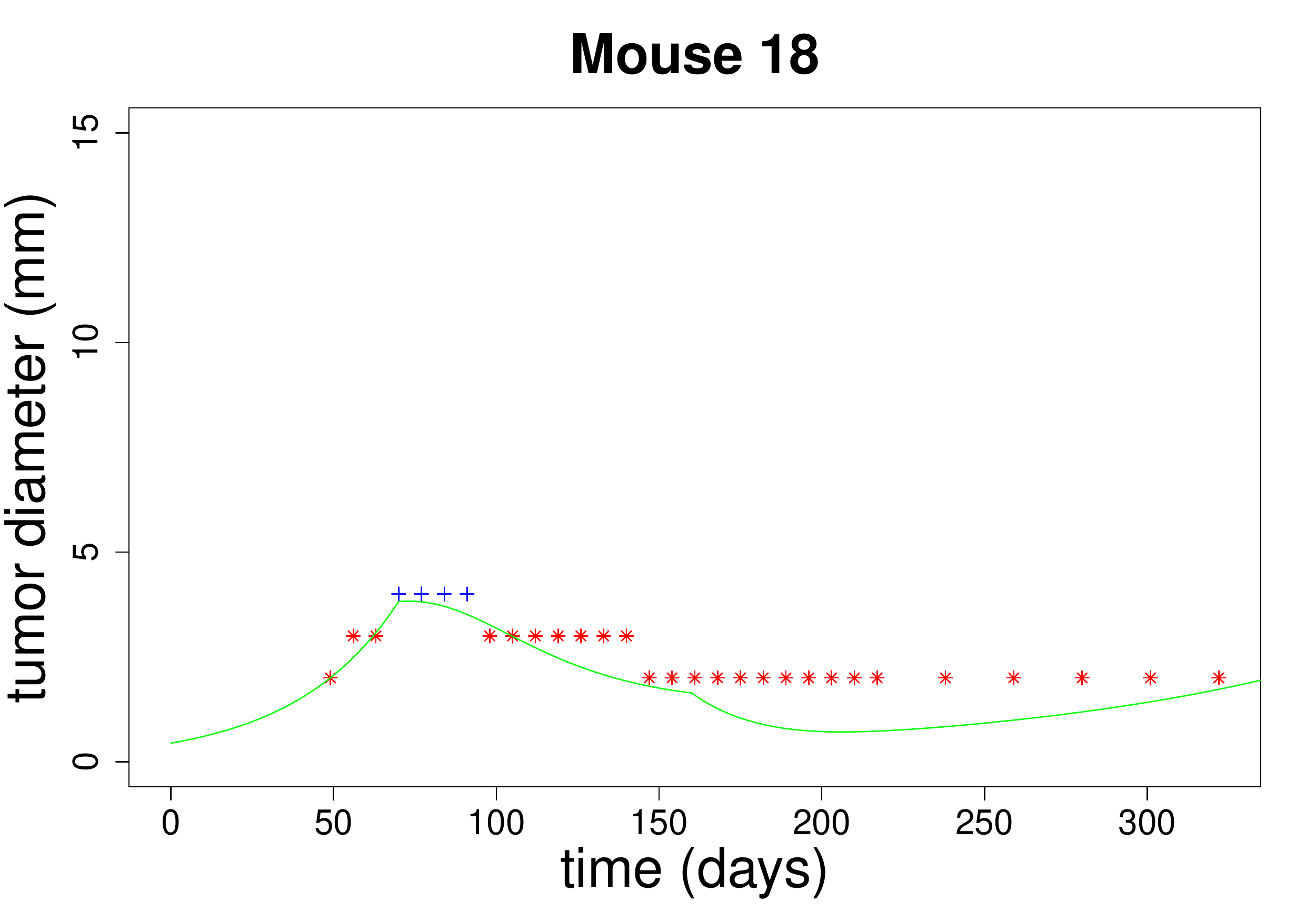}} 

     \caption{\scriptsize{Individual fits for mice in CTRL group (a),  ACT group (b), and  ACT+Re group (c,d). Blue + marks represent no censored observations,   red asterisk   the censored observations,   green line the fitted tumor size dynamics.}}
     \label{fits_indiv_final} 
\end{figure}

\subsection{{Good fits for the three groups of mice}}

As shown in Figures \ref{fits_indiv_final} and \ref{fits_all_inds_final}, data are well fitted, the growth and treatment phases are well enough captured despite the presence of censored data.
The tumor growth phase is well fitted for CTRL mice as well as for treated mice (before the beginning of therapy). Treatment effect in the tumor size dynamics from the 70th day is also well highlighted in both ACT and ACT+Re groups. Furthermore, the restimulation effect is well marked in the tumor size dynamics for ACT+Re mice.

\subsection{Biological relevance of {the additional effect due to treatment $s_A$}}

As cancer cells escape therapy, it is expected that the additional switching rate $s_A$ from differentiated to dedifferentiated melanoma cells is higher than the killing rate $d_{MT}$ of differentiated cells by T cells \cite{Meri}. Indeed, both mechanisms appear as quadratic terms involving the number of differentiated melanoma cells. Thus, if the cytokines and T cells populations sizes are similar,   rates $s_A$ or $d_{MT}$ determine which mechanism prevails. 

The estimated parameters confirm this relation: $s_A > d_{MT}$, as shown in Table \ref{table:tableResu}. This is satisfactory concerning the biological modeling of the escape mechanism. Note that the calibration of model parameters in \cite{Loren} led on the contrary to $d_{MT}$ much greater than $s_A$. However, their results were calibrated manually without any comparison to real biological data. This reflects the relevance of   our statistical approach.  

\subsection{Very small T cell exhaustion probability for population parameters}

\label{ssec:ISResu}
The T cell exhaustion probability estimated by the IS algorithm is very close to zero  }for   individual and population parameters in both ACT and ACT+Re groups. Figure \ref{sim_Rcpp_ACT_exhaustion_and_survive} shows the dynamics of cancer cells, T cells, and cytokines for population parameters in the ACT group. For these   parameter values, T cells thus survive and continue to kill differentiated cells {\color{black}(whose number is controlled)}. However, the therapy enhances the switch towards dedifferentiated melanoma cells, which are not killed by T cells and thus grow exponentially.
{The tumor size regrows with a large ratio of dedifferentiated over differentiated melanoma cells. This is in accordance with Figure 1c of \cite{Landsberg}.}
%
%
\begin{figure}
\centering 	\centering\includegraphics[height = 0.6\textwidth, width=0.99\textwidth]
{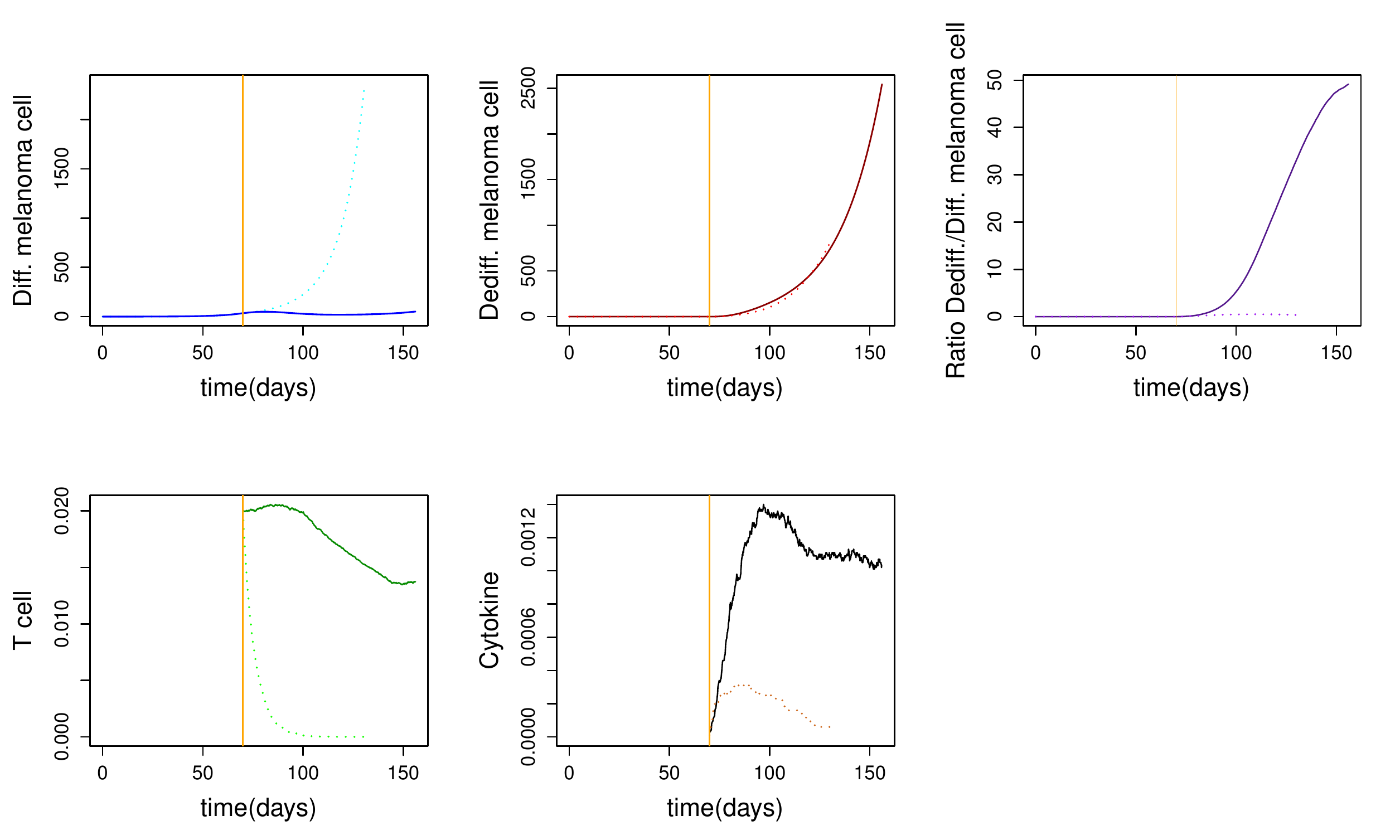}
	\caption{\scriptsize{Rescaled dynamics (population size/$K$) of differentiated cancer cells (top-left), dedifferentiated cancer cells (top-center), ratio of dedifferentiated   over differentiated cells (top-right), T cells (bottom-left), and cytokines (bottom-center). Scenarios: T cell survive (solid lines,   population parameters of   ACT group); T cell exhaustion (dotted lines,   population parameters of   ACT group and $d_T = $ quantile of order 99\% of $d_{T_i}$). Vertical orange line: treatment start.}} 
	\label{sim_Rcpp_ACT_exhaustion_and_survive}
\end{figure}

\subsection{Evolution of T cell exhaustion probability with respect to therapy parameters}

Some of the estimated parameters come with a non negligible random effect. We study the T cell exhaustion by naturally taking into account their distribution.  We write $q_{x}(\mu)$ for the quantile of order $x\%$ of the parameter $\mu$.

\begin{figure}
\centering 	\centering\includegraphics[height = 0.4\textwidth, width=0.7\textwidth]{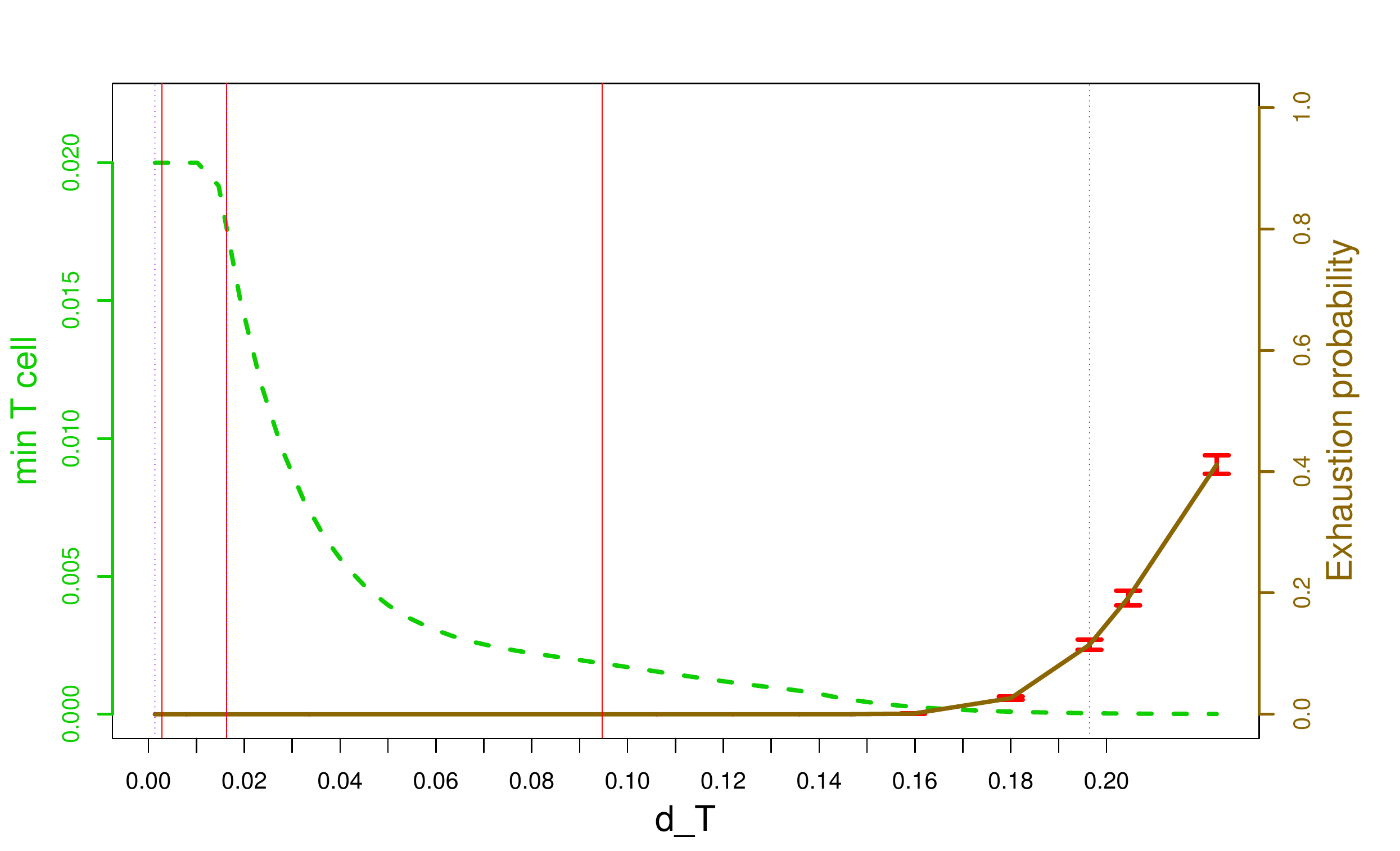}
	\caption{\scriptsize{T cell exhaustion probability (brown solid line) and   global minimum of   T cell population size (green dashed line) according to  $d_T$  in ACT group (other parameters   fixed to their estimated population mean). Black curve: T cell exhaustion probability as a function of $d_T$. 
	Red solid vertical lines: quantiles of order 5\%, 50\%, 95\% of   $d_T$ distribution; purple  dotted vertical lines  quantiles of order  1\% and 99\%. }}
	\label{probaExhaustion_and_MinTc_vs_dT}
\end{figure}

Death rate of T cells $d_T$   is one of the most relevant therapy parameters. We thus estimate the T cell exhaustion probability as a function of $d_T$ in its estimation range {\color{black}($d_T \in [0,q_{99}(d_T)]$)} when the other parameters are fixed to their estimated population values   (still in ACT group). 
  Figure \ref{probaExhaustion_and_MinTc_vs_dT} shows that the T cell exhaustion probability increases with $d_T$, and even reaches non negligible values. This corroborates the conjecture in \cite{Loren} asserting that there exist biological parameters for which the probability of relapse due to T cell exhaustion is non negligible. Note that when T cell exhaustion occurs, differentiated melanoma cells grow exponentially fast again, and become the most numerous in the final state of the tumor.

Figure \ref{probaExhaustion_and_MinTc_vs_dT} gives  the probability of T cell exhaustion as a function of parameters with random effects (the others parameters being fixed to their population mean).  We observe that the first minimum of the deterministic T cell trajectory becomes deeper and deeper as the parameters $d_T$, $d_{MT}$ or $l_{A}^{\text{prod}}$ increase. On the other hand, the minimum {increases} when disease parameters $b_M$, $b_D$ or $n_{M_0}$ increase. 
 Furthermore, the minimum increases also with the cytokine clearance rate $d_A$. 
Given these observations, we can expect that the joint variation of these parameters  will lead us to interesting therapeutic scenarios.

{\color{black}

\subsection{Treatment optimization to minimize the probability of relapse due to T cell exhaustion}

To reduce the number of parameter values of $\{d_{MT}, \ d_T, \ l_{A}^{\text{prod}}, \ n_{M_0}\}$ to study, we focus on   combinations of quantile values 
leading to an increased risk of T cell exhaustion. 
For that purpose, we define a threshold on the value of the function $g$ (giving the value of the first minimum of the deterministic T cell trajectory) below which stochastic fluctuations have a good chance to lead to the T cell exhaustion. 
By setting this threshold to $S_1 = 7 \times 10^{-5}$ (first  threshold of IS algorithm), we identify $9$ sets of parameter values of $\{d_{MT}, \ d_T, \ l_{A}^{\text{prod}}, \ n_{M_0}\}$ for which the deterministic dynamics of  the T cell   population crosses the exhaustion threshold $S_1$.
Table \ref{table:tableOptim_quantiles_5_50_95} presents the optimization results for these $9$ sets of parameter values presenting an increased risk of T cell exhaustion. Column ACT corresponds to   optimization criterion \eqref{optimizationACT} and   column ACT+Re to   criterion \eqref{optimizationACTRE}. 
 Figures \ref{optACT1} and \ref{optACT2} show optimal treatment parameters corresponding to two sets of parameter values for the optimization criterion \eqref{optimizationACT}  (Figure \ref{optACT1} presents a risk of T cell exhaustion, Figure \ref{optACT2}   without an exhaustion risk). Figures \ref{optACTRE1} and \ref{optACTRE2} show the optimization results for the same two sets of parameters but for   criterion \eqref{optimizationACTRE}. 
Note that parameter values leading to an increased risk of T cell exhaustion mainly correspond to high values of parameters $d_{T}$ and $l_{A}^{\text{prod}}$ as shown for their quantile of order 95\% in {\color{black}Figure \ref{minTc_vs_dose70_ACT_and_vs_dose70_tRe_dRe_ACTRE_group}}. 

\begin{figure}
     \centering
     \subcaptionbox{{\tiny ACT: $\text{q}_{95}(d_{MT})$, $\text{q}_{95}(d_T)$, $\text{q}_{95}(l_{A}^{\text{prod}})$} \label{optACT1}}{\includegraphics[width=0.4\textwidth]{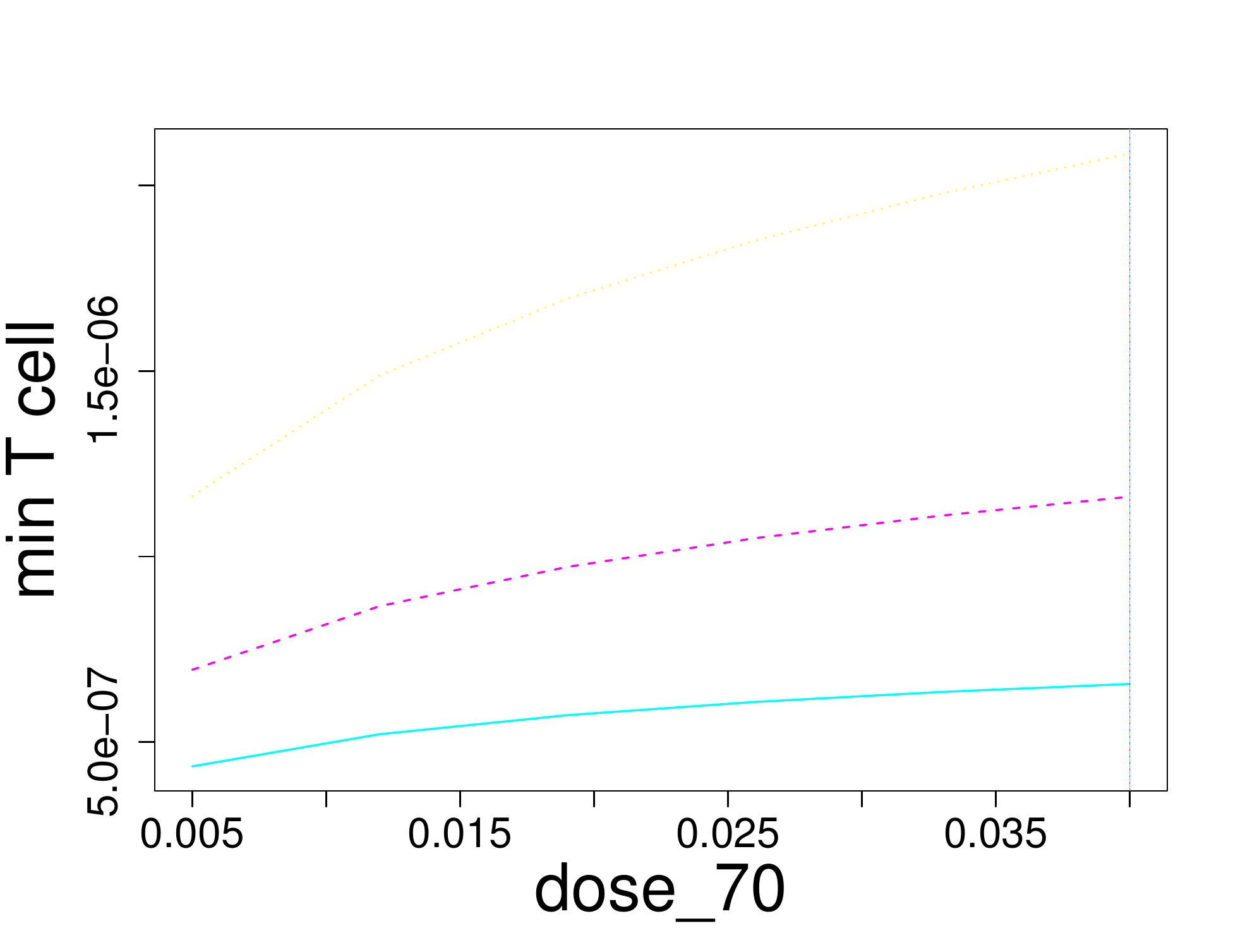}}\quad
     \subcaptionbox{{\tiny ACT: $\text{q}_{5}(d_{MT})$, $\text{q}_{95}(d_T)$, $\text{q}_{50}(l_{A}^{\text{prod}})$\label{optACT2}}}{\includegraphics[width=0.4\textwidth]{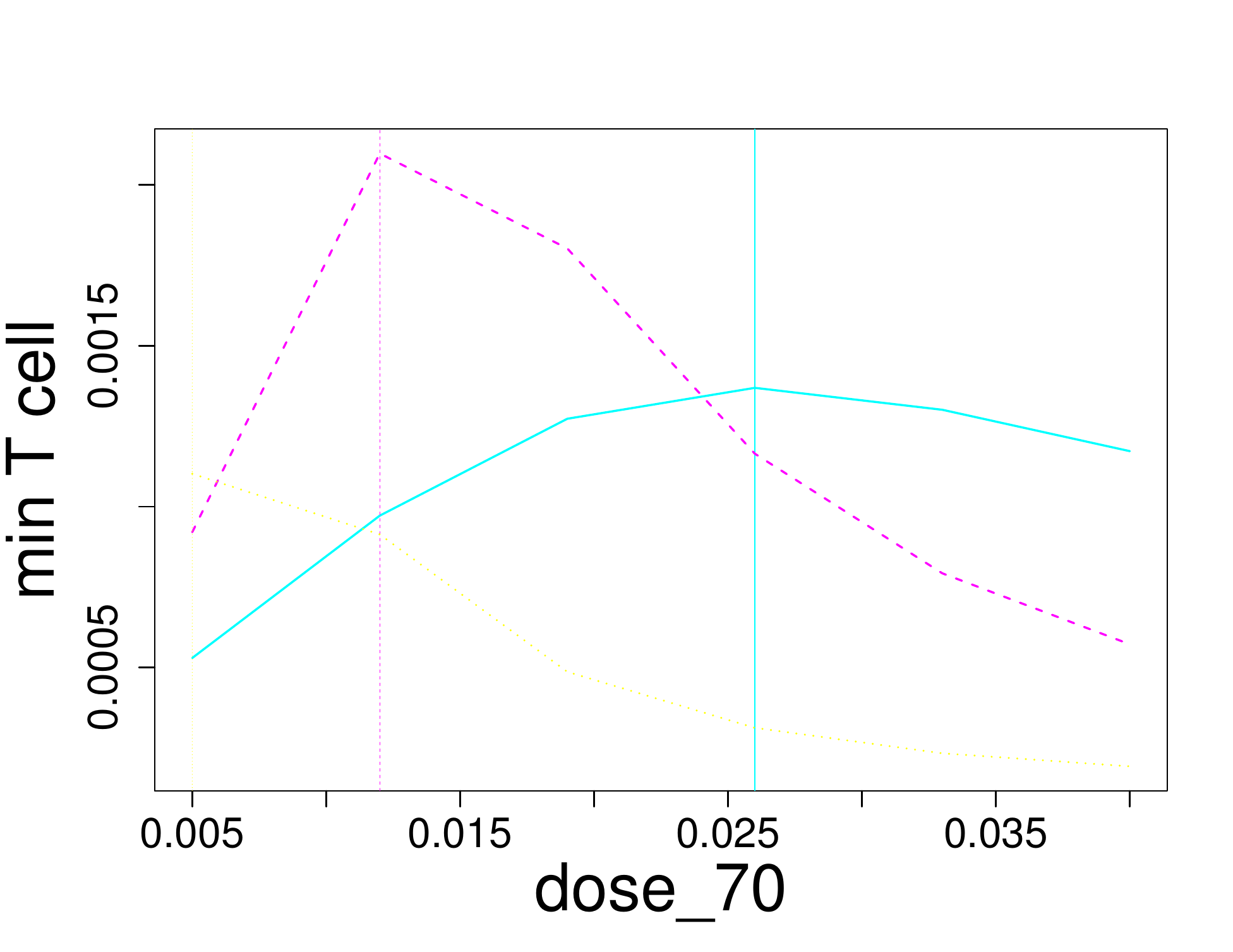}}\\
     \subcaptionbox{{\tiny  ACT+Re: $d_ {70}^{opt_{**}} =  0.04$, $\text{q}_{95} (d_{MT})$, $\text{q}_{95}(d_T)$, $\text{q}_{95}(l_{A}^{\text{prod}})$, $\text{q}_{95}(n_{M_{0}})$ \label{optACTRE1}}}{\includegraphics[width=0.4\textwidth]{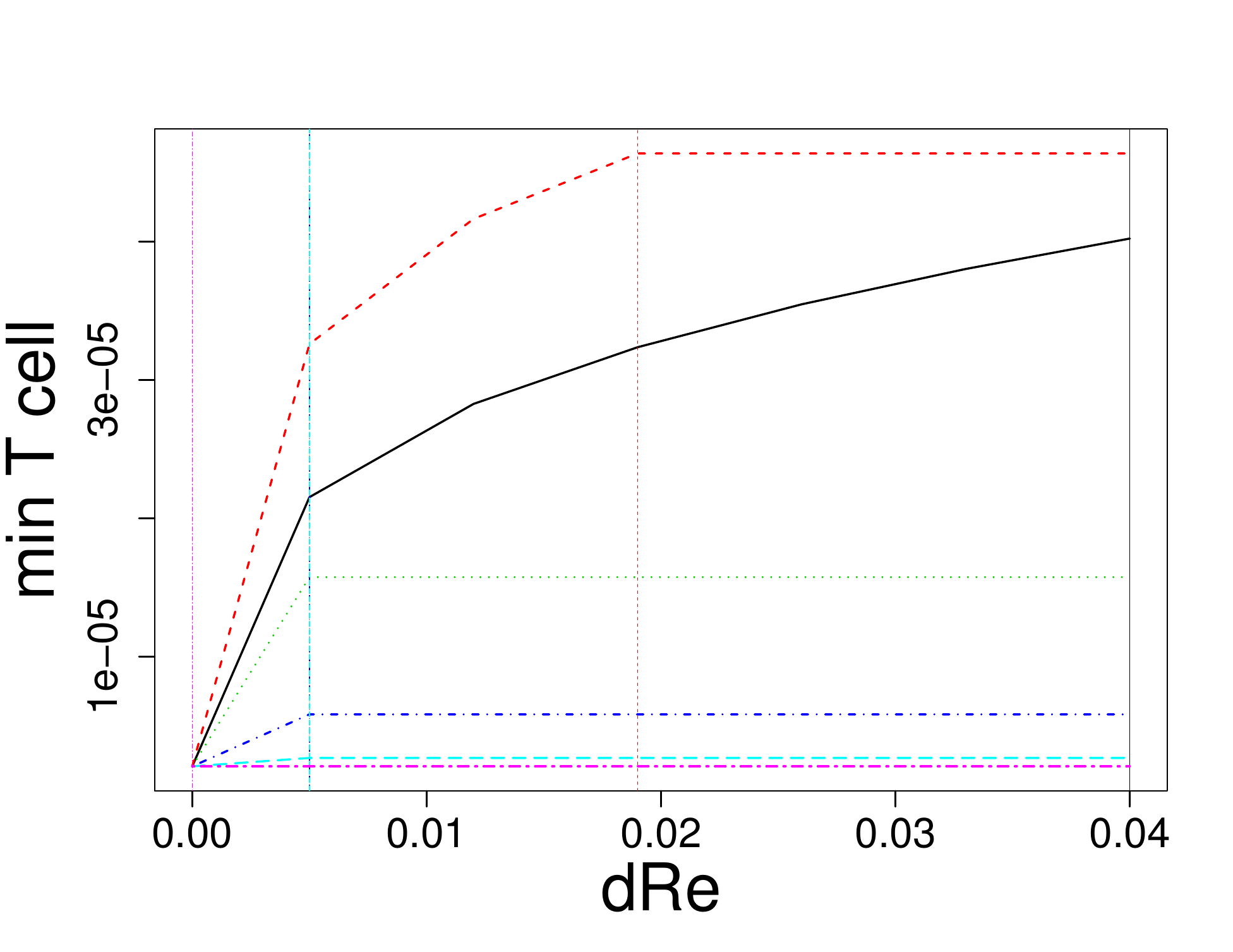}}\quad
     \subcaptionbox{{\tiny ACT+Re: $d_{70}^{opt_{**}} = 0.005$, $\text{q}_{5}(d_{MT})$, $\text{q}_{95}(d_T)$, $\text{q}_{50}(l_{A}^{\text{prod}})$, $\text{q}_{95}(n_{M_{0}})$ \label{optACTRE2}}}{\includegraphics[width=0.4\textwidth]{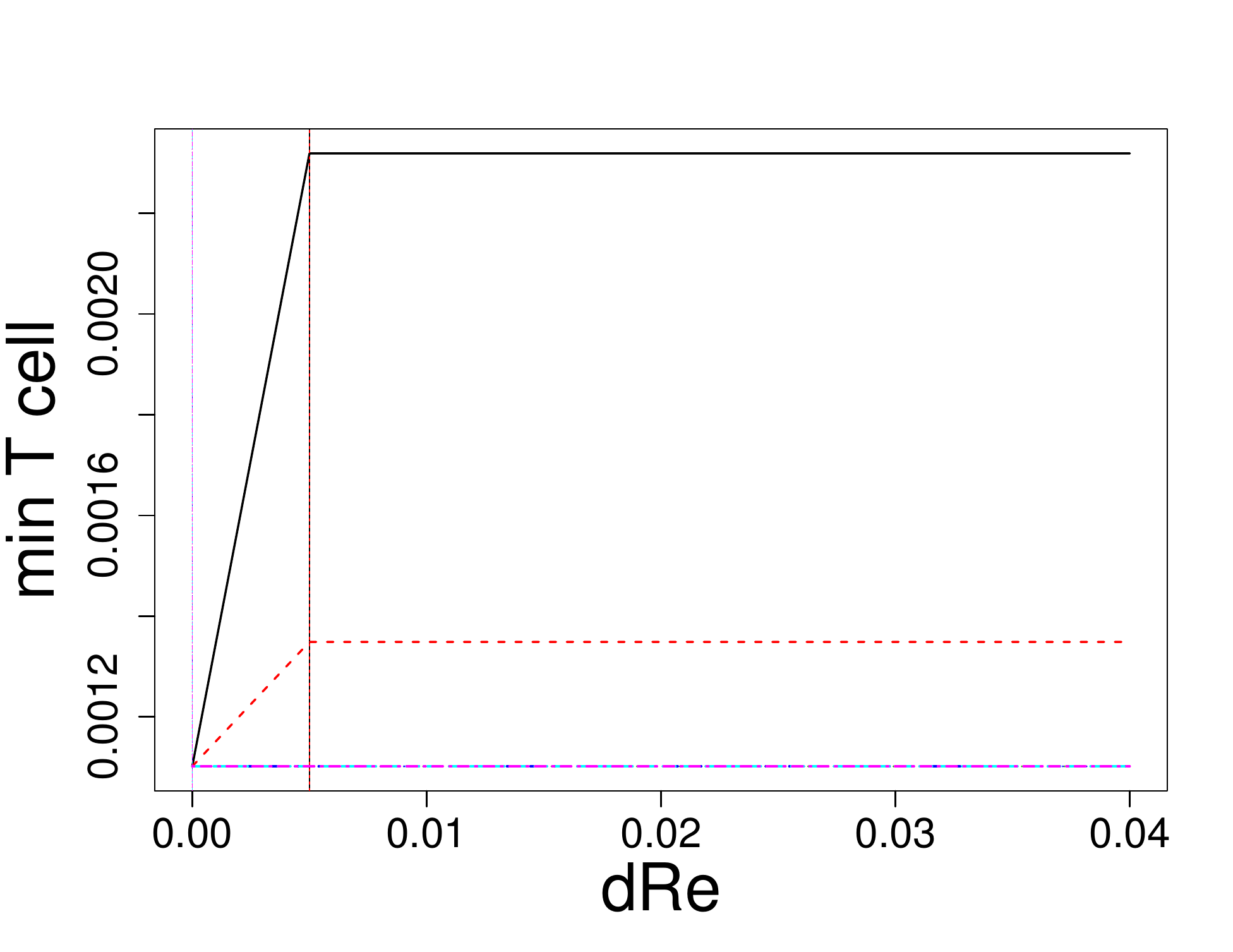}}

    \caption{\scriptsize{Global minimum of  T cell population size according to treatment parameters using quantiles of order  $\{5, 50, 95\}\%$, for  $ d_{MT}$, $d_{T}$, $l_{A}^{\text{prod}}$, $n_{M_{0}}$.  In ACT group (a,b): solid curve in cyan ($\text{q}_{5}(n_{M_{0}})$), dashed curve in purple ($\text{q}_{50}(n_{M_{0}})$), dotted curve in yellow ($\text{q}_{95}(n_{M_{0}})$). Vertical lines: optimal doses $d_{70}^{\text{opt}_{*}}$ for the corresponding $n_{M_{0}}$.  
In ACT+Re group (c,d):   curves for   discrete retreatment times $t_{\text{Re}}=130$  (black solid), 142 (red dashed), 154 (green dotted), 166  (blue dotdashed), 178 (cyan longdashed), 190 (purple twodashed). Vertical lines: optimal $d_{\text{Re}}^{\text{opt}}$. }}
     \label{minTc_vs_dose70_ACT_and_vs_dose70_tRe_dRe_ACTRE_group}
\end{figure}

{\footnotesize
\begin{table}[!h] 
\centering
\begin{tabular}{lllllllll}
\hline
$d_{MT}$,  $d_T$,  $l_{A}^{\text {prod}}$, $n_{M_0}$ & par. & {\scriptsize ACT} & $g_{\text{ACT}}^{\text{opt}}$ & $g_{\text{ACT}}^{\text{initial}}$ & {\scriptsize ACT+Re} & $ g_{\text{ACT+Re}}^{\text{opt}} $ & $g_{\text{ACT+Re}}^{\text{initial}} $ \\ \hline
\multirow{3}{*}{$\text{q}_{5}$, $\text{q}_{95}$, $\text{q}_{95}$, $\text{q}_{5}$} & $d_{70}^{\text{opt}}$ & $0.04$ & $8.29\times 10^{-7}$& $6.77\times 10^{-7}$ & $0.019$  & $1.92 \times 10^{-5}$ & $4.45  \times 10^{-6}$  \\
 & $d_{\text{Re}}^{\text{opt}}$ & - & & & $0.04$ \\
 & $t_{\text{Re}}^{\text{opt}}$ & - & & &  $142$ \\  \hline
 \multirow{3}{*}{$\text{q}_{5}$, $\text{q}_{95}$, $\text{q}_{95}$, $\text{q}_{50}$} & $d_{70}^{\text{opt}}$ & $0.04$ & $1.36\times 10^{-6}$ & $1.10\times 10^{-6}$ &$0.033$  & $3.49  \times 10^{-5}$ & $4.73 \times 10^{-6}$  \\
 & $d_{\text{Re}}^{\text{opt}}$ & - & & & $0.04$ \\
 & $t_{\text{Re}}^{\text{opt}}$ & - & & &  $142$ \\  \hline
 \multirow{3}{*}{$\text{q}_{5}$, $\text{q}_{95}$, $\text{q}_{95}$, $\text{q}_{95}$} & $d_{70}^{\text{opt}}$ & $0.04$ & $2.33\times 10^{-6}$ & $1.86\times 10^{-6}$ & $0.04$  & $4.66  \times 10^{-5}$ & $5.22 \times 10^{-6}$  \\
 & $d_{\text{Re}}^{\text{opt}}$ & - & & & $0.012$ \\
 & $t_{\text{Re}}^{\text{opt}}$ & - &  & & $142$ \\  \hline
 \multirow{3}{*}{$\text{q}_{50}$, $\text{q}_{95}$, $\text{q}_{95}$, $\text{q}_{5}$} & $d_{70}^{\text{opt}}$ & $0.04$ & $7.99\times 10^{-7}$ & $6.61\times 10^{-7}$ & $0.019$  & $1.87 \times 10^{-5}$ & $4.44 \times 10^{-6}$  \\
 & $d_{\text{Re}}^{\text{opt}}$ & - & & & $0.04$ \\
 & $t_{\text{Re}}^{\text{opt}}$ & - & & & $142$ \\  \hline
 \multirow{3}{*}{$\text{q}_{50}$, $\text{q}_{95}$, $\text{q}_{95}$, $\text{q}_{50}$} & $d_{70}^{\text{opt}}$ & $0.04$ &$1.33\times 10^{-6}$ & $1.08\times 10^{-6}$ & $0.033$ & $3.40 \times 10^{-5} $ & $4.72 \times 10^{-6}$  \\
 & $d_{\text{Re}}^{\text{opt}}$ & - & & & $0.04$ \\
 & $t_{\text{Re}}^{\text{opt}}$ & - & & & $142$ \\  \hline
 \multirow{3}{*}{$\text{q}_{50}$, $\text{q}_{95}$, $\text{q}_{95}$, $\text{q}_{95}$} & $d_{70}^{\text{opt}}$ & $0.04$ &$2.29\times 10^{-6}$ & $1.83\times 10^{-6}$ & $0.04$  & $4.66 \times 10^{-5} $  & $5.20 \times 10^{-6}$ \\
 & $d_{\text{Re}}^{\text{opt}}$ & - & & & $0.019$ \\
 & $t_{\text{Re}}^{\text{opt}}$ & - &  & & $142$ \\  \hline
 \multirow{3}{*}{$\text{q}_{95}$, $\text{q}_{95}$, $\text{q}_{95}$, $\text{q}_{5}$} & $d_{70}^{\text{opt}}$ & $0.04$ & $6.56\times 10^{-7}$ & $5.78\times 10^{-7}$  &$0.019$  & $1.64 \times 10^{-5} $ & $4.38 \times 10^{-6}$ \\
 & $d_{\text{Re}}^{\text{opt}}$ & - & & & $0.04$ \\
 & $t_{\text{Re}}^{\text{opt}}$ & - &  & & $142$ \\  \hline
 \multirow{3}{*}{$\text{q}_{95}$, $\text{q}_{95}$, $\text{q}_{95}$, $\text{q}_{50}$} & $d_{70}^{\text{opt}}$ & $0.04$ & $1.16\times 10^{-6}$ & $9.84\times 10^{-7}$  &$0.033$   &$2.99 \times 10^{-5}$  & $4.65 \times 10^{-6}$ \\
 & $d_{\text{Re}}^{\text{opt}}$ & - & & & $0.04$ \\
 & $t_{\text{Re}}^{\text{opt}}$ & - & & & $142$ \\  \hline
 \multirow{3}{*}{$\text{q}_{95}$, $\text{q}_{95}$, $\text{q}_{95}$, $\text{q}_{95}$} & $d_{70}^{\text{opt}}$ & $0.04$ & $2.09\times 10^{-6}$ &  $1.72\times 10^{-6}$ &$0.04$ & $4.63 \times 10^{-5}$  & $5.13 \times 10^{-6}$ \\
 & $d_{\text{Re}}^{\text{opt}}$ & - & & & $0.019$ \\
 & $t_{\text{Re}}^{\text{opt}}$ & - &  & & $142$ \\  \hline

\end{tabular}
\caption{\scriptsize{Optimization results with quantiles $q_{\alpha}(\mu)$, $\alpha \in \{5, 50, 95\}\%$, and $\mu\in\{d_{MT},d_{T},l_{A}^{\text{prod}},n_{M_{0}}\}$ in   ACT and ACT+Re groups. Column $g_{\text{ACT}}^{\text{opt}}$ corresponds to $g(d_{70}^{\text{opt}_{*}})$, column $g_{\text{ACT}}^{\text{initial}}$ to $g(d^{\text{initial}})$, column $g_{\text{ACT+Re}}^{\text{opt}}$   to $g(d_{70}^{\text{opt}_{**}}, d_{\text{Re}}^{\text{opt}},t_{\text{Re}}^{\text{opt}})$ and $g_{\text{ACT+Re}}^{\text{initial}}$ to $g(d^{\text{initial}}, d^{\text{initial}},t_{\text{Re}}^{\text{initial}})$. Doses $d_{70}, \ d_{\text{Re}}$   in number of T cells/$K$,   restimulation time $t_{\text{Re}}$ in days.}}
\label{table:tableOptim_quantiles_5_50_95}
\end{table}
}

\label{sec:ResuOptVal}

\section{Discussion on optimized doses and restimulation times}

\subsection{{\color{black}Higher treatment doses and earlier restimulation times are not always {better}}}
Using  quantiles of order 5\%, 50\%, 95\% of $d_{MT}$, $d_{T}$, $l_{A}^{\text{prod}}$, and $n_{M_{0}}$, we observe different therapeutic scenarios. For most of   parameter values, the optimal doses are the maximal ones: $d_{70}^{\text{opt}}$, $d_{\text{Re}}^{\text{opt}} = d_{\text{max}}$ (see Figure \ref{optACT1}).
For  other sets of quantile values (not presented in this paper), the optimal retreatment time is the minimal one $t_{\text{Re}}^{\text{opt}} = t_{\text{min}}^{\text{Re}}$. 
We can thus expect that higher treatment doses would further delay the T cell exhaustion or that an earlier restimulation would reduce the risk of an early exhaustion.
However, the treatment parameters are optimized over optimization intervals which take into account biological considerations as explained in Section \ref{sec:OptValComput}. 
Particularly, the Cytokine Release Syndrome (one of the most common, and potentially life threatening side effects of immunotherapy) is associated with increased doses of immunotherapy \cite{maude2014managing}. 
It may thus be experimentally non trivial to go beyond those intervals. 

For most rate parameters (Figure \ref{Tcell_VS_DM_tT_50_dT_95_Lw_50_M0_50_pour_doses_croissantes} in Appendix), the larger dose, the deeper the T cell global minimum, therefore the higher the probability of T cell exhaustion.
For some parameter values, the optimal restimulation dose is zero: $d_{\text{Re}}^{\text{opt}} = 0$. In these cases, the treatment dose $d_{70}$ is sufficient to avoid exhaustion.
Finally, several parameter values (Figures \ref{optACT2}, \ref{optACTRE1}, \ref{minTc_vs_dose70_ACT_tT_95}, \ref{minTc_vs_dose70_tRe_dRe_ACTRE_quantiles_5_50_95_interessant2}, \ref{minTc_vs_dose70_tRe_dRe_ACTRE_quantiles_5_50_95_interessant}) led to optimal treatment doses and restimulation times which differ from the tested experimental ones. 

\subsection{{Treatment optimization leads to values of $g$ which are higher in the ACT+Re group than in the ACT group}}
 Optimization benefit is evaluated by   the probability of T cell exhaustion using the IS algorithm. As the function $g$ is related to the exhaustion probability,  we compare the values of $g$ using respectively the initial treatment parameters (initial doses, and restimulation time provided in Section  \ref{sec:OptValComput}, and the optimized treatment parameters (Section \ref{sec:ResuOptVal}).
More precisely, let {$g_{\text{G}}^{\text{initial}}$ and $g_{\text{G}}^{\text{opt}}$ (for $G \in \{\text{ACT}, \text{ACT+Re}\}$) represent the values of $g_{\text{G}}$ computed respectively using the initial treatment parameters and the optimized treatment parameters.} Let $\text{Imp}_{\text{ACT+Re}} = ({g_{\text{ACT+Re}}^{\text{opt}} - g_{\text{ACT+Re}}^{\text{initial}}})/{g_{\text{ACT+Re}}^{\text{opt}}}$
and $\text{Imp}_{\text{ACT}} = ({g_{\text{ACT}}^{\text{opt}} - g_{\text{ACT}}^{\text{initial}}})/{g_{\text{ACT}}^{\text{opt}} }$ denote the relative shifts of $g$ in the ACT+Re (resp. ACT) groups after treatment optimization, {for the same values of the parameters}. 
The two shifts are positive. Moreover, the difference $\text{Imp}_{\text{ACT+Re}} - \text{Imp}_{\text{ACT}}$ is always positive with a mean value equal to {${(65.7 \pm 4.8) \%}$} 
for the nine sets of parameters presenting an increased risk of T cell exhaustion (Table \ref{table:tableOptim_quantiles_5_50_95}). Thus, the best improvements are noted in ACT+Re mice compared to ACT mice. This   makes sense since ACT+Re mice have benefited from restimulation, which boosts the effects of optimization.

Note however that the treatment optimization only minimizes the risk of T cell exhaustion, the risk does not vanish completely   as seen in Table \ref{table:tableOptim_quantiles_5_50_95}: $g^{\text{opt}}$ is still smaller than $S_1$. Considering wider optimization intervals could improve the results provided that the new limits remain biologically reasonable.

\subsection{The larger $n_{M_0}$, the faster the tumor growth even for higher treatment doses}
 It has been biologically observed \cite{Meri} that when the tumor size exceeds a certain threshold at the beginning of the treatment ($t = 70$), the tumor control is less efficient even if more T cells are stimulated. Figure \ref{tumorSize_vs_time} (Appendix) representing   tumor size dynamics for different values of $n_{M_0}$ supports this observation: the larger $n_{M_0}$, the larger tumor size at $t = 70$, the faster the tumor growth (even for $d_{70}  = d_{70}^{\text{opt}_{*}} = d_{\text{max}}$). Large treatment doses $d_{70}$ do not slow down much the evolution of the tumor size for large initial value of differentiated melanoma cells. 


\section{Conclusion}
 
In cancer treatment, resistance to therapy is a major problem. Understanding the mechanism of this resistance is then crucial. Parameter estimation for   mathematical       melanoma cancer immunotherapy model \cite{Loren} using biological data from \cite{Landsberg} was essential in achieving this goal. Indeed,  the estimated parameters lead to a statistically satisfactory model, which replaces the empirical calibration approach of Baar et al.\ \cite{Loren}, and which constitutes a first necessary step for quantitative studies. We use these parameters to simulate  realistic stochastic phenomena arising in the therapy and to highlight the existence of sets of biological parameters leading to each of the types of relapses identified by  \cite{Loren}. Moreover, we are able to quantitatively estimate the probability of relapse due to  T cell exhaustion, which allows to evaluate the quality of the treatment. 

In addition, we confirme two important conjectures from the authors of  \cite{Landsberg}. 
First, the therapy escape mechanism suggests that the switch rate $s_A$ induced by cytokines should be higher than the rate $d_{MT}$ at which T cells kill differentiated melanoma cells. 
Second, biological observations \cite{Meri} show that when the tumor size exceeds a certain threshold at the beginning of the treatment, the tumor control is less efficient even if more T cells are stimulated. We indeed highlight the low effect of high treatment doses $d_{70}$ when the tumor size reaches a certain level at the beginning of the treatment.

With the estimated parameters, we optimize the treatment protocol using the deterministic system  by plotting  the value of the first local minimum of the T cell population as a function of treatment parameters. By linking the value of the minimum to the exhaustion probability of T cells, we provide optimal treatment doses and (re)stimulation time(s) which minimize the probability of relapse due to T cell exhaustion.

An alternative to  the deterministic system would be  a diffusion approximation of the stochastic model. A stochastic aspect would thus be included in the problem while keeping reasonable computational costs. An interesting statistical challenge, which is a work in progress,  is to perform the parameter estimation in this stochastic diffusion approximation. The difficulties are mainly due to the predator-prey setting and the  high dimensionality of the problem. 

This paper focuses on the relapse mechanism of melanoma which is mediated by the T cell exhaustion. As explained in the introduction, we use a model which simplifies a lot   the real mechanisms underlying the exhaustion process. Certain mechanisms must still be studied for a better understanding of the resistance to the therapy. This includes, for example, the role of IFN-$\gamma$, or the relapse due to dedifferentiated melanoma cells not killed by T cells. There are experimental prospects (requiring modifications of the treatment protocol) to handle the latter problem. One solution may be to stimulate another type of T cells which can kill the dedifferentiated melanoma (see \cite{Loren} for a theoretical study of this treatment protocol).  Another solution is to find a way to control the switching rate $s_A$ of differentiated melanoma to dedifferentiated melanoma cells, as recently considered by Glodde, Kraut et al.\ \cite{Glodde-Kraut}. 
Indeed, if $s_A$ is well controlled in the therapy (allowing the number of dedifferentiated cells to be driven to low enough values), then the problem of complete healing will come down to the control of T cell exhaustion. Thus, our work is a useful tool for new  protocols of tumor treatment.}

\bibliographystyle{abbrv}
\bibliography{biblio}


\addresseshere

\clearpage
\newpage

 \renewcommand\thesection{\Alph{section}}
\setcounter{section}{0}
\renewcommand{\theequation}{\Alph{section}.\arabic{equation}}

 \section{Appendix}

The code used to simulate the stochastic dynamics has been written in C++ and ran under the software R \cite{Rsoftware} through the package Rcpp \cite{eddelbuettel2011rcpp}. 
\begin{table}[H] 
\centering
{\footnotesize
\begin{tabular}{llll}
  \hline
  {\small Parameter (1/days)} & Interpretation   \\
  \hline
   \vspace{0.05cm}
   $b_{M}$  &  Difference between division and death rates of differentiated melanoma cells \\
    \vspace{0.05cm}
   $b_{D}$ & Difference between division and death rates of dedifferentiated melanoma cells \\
   
    \vspace{0.05cm}
   $s_{MD}$ & Natural switch rate from differentiated to dedifferentiated melanoma cells  \\
    \vspace{0.05cm}
   $s_{DM} $ & Natural switch rate from dedifferentiated to differentiated melanoma cells    \\
   
    \vspace{0.05cm}
   $b_T$ & Division rate of T cells \\
     \vspace{0.05cm}
   $d_{T}$ & Exhaustion rate of T cells  (interpreted as death rate of active T cells) \\
    \vspace{0.05cm}
   $d_{MT}$ & Therapy induced death rate of differentiated melanoma cells  \\

   $s_{A}$ &  Therapy induced switch rate from differentiated to dedifferentiated melanoma cells \\

 \vspace{0.055cm}
   $d_{A}$ & Clearance rate of cytokines $TNF_\alpha$ \\
   
   $l_{A}^{\text{prod}}$ &  Secretion rate of cytokines $TNF_\alpha$ \\ 
   
%
%
%
%
      
  \hline
  \end{tabular}
  }
 \caption[Table of  melanoma model parameters]{\small{{Table of      melanoma model parameters.}}}  
  \label{table:tableParamModeleImmuno}
  \end{table}
%
%
%

\begin{figure}
    \includegraphics[width=.22\textwidth]{Images/LRT/Final/ind_fits_1.pdf}\hfill
    \includegraphics[width=.22\textwidth]{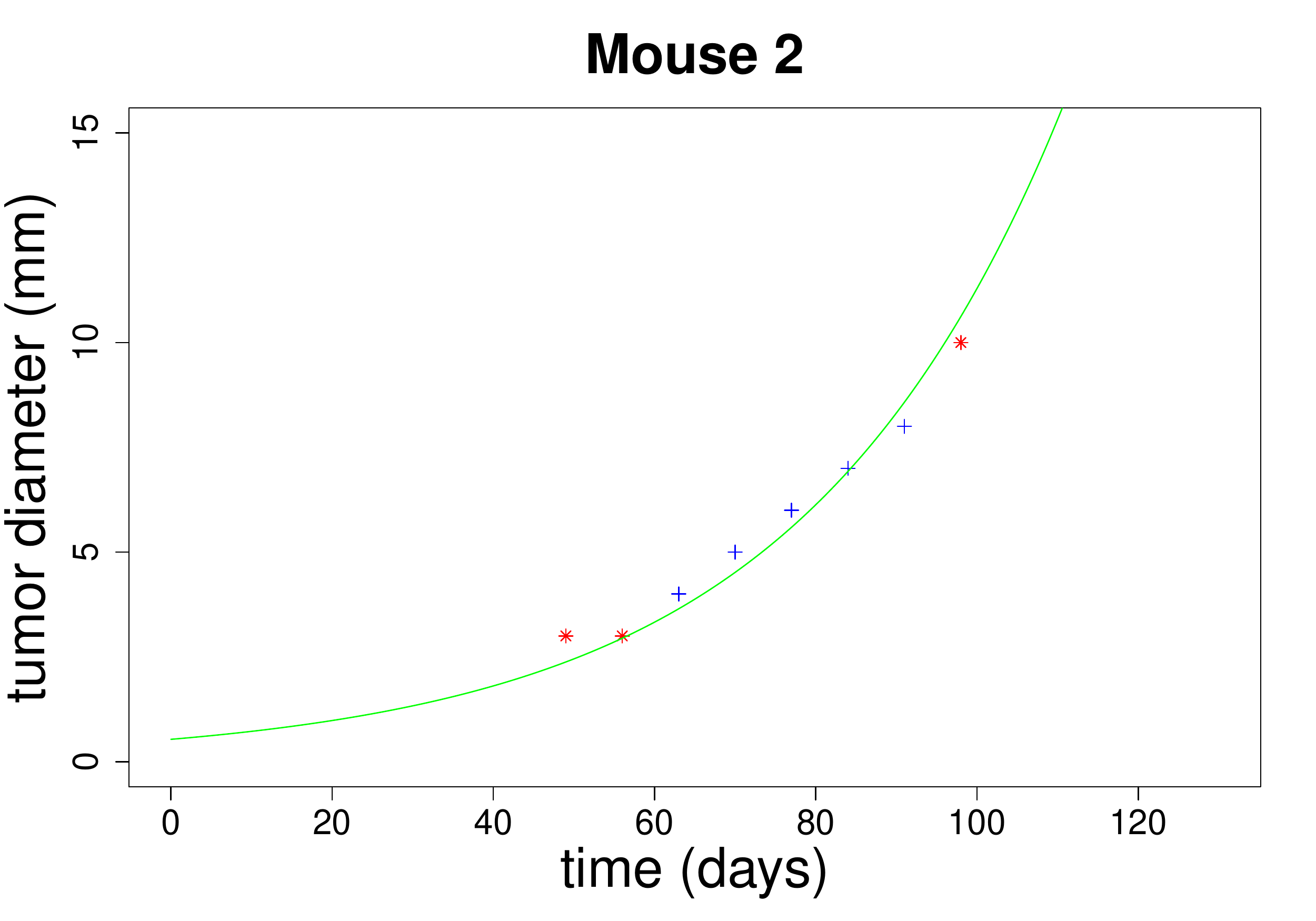}\hfill
    \includegraphics[width=.22\textwidth]{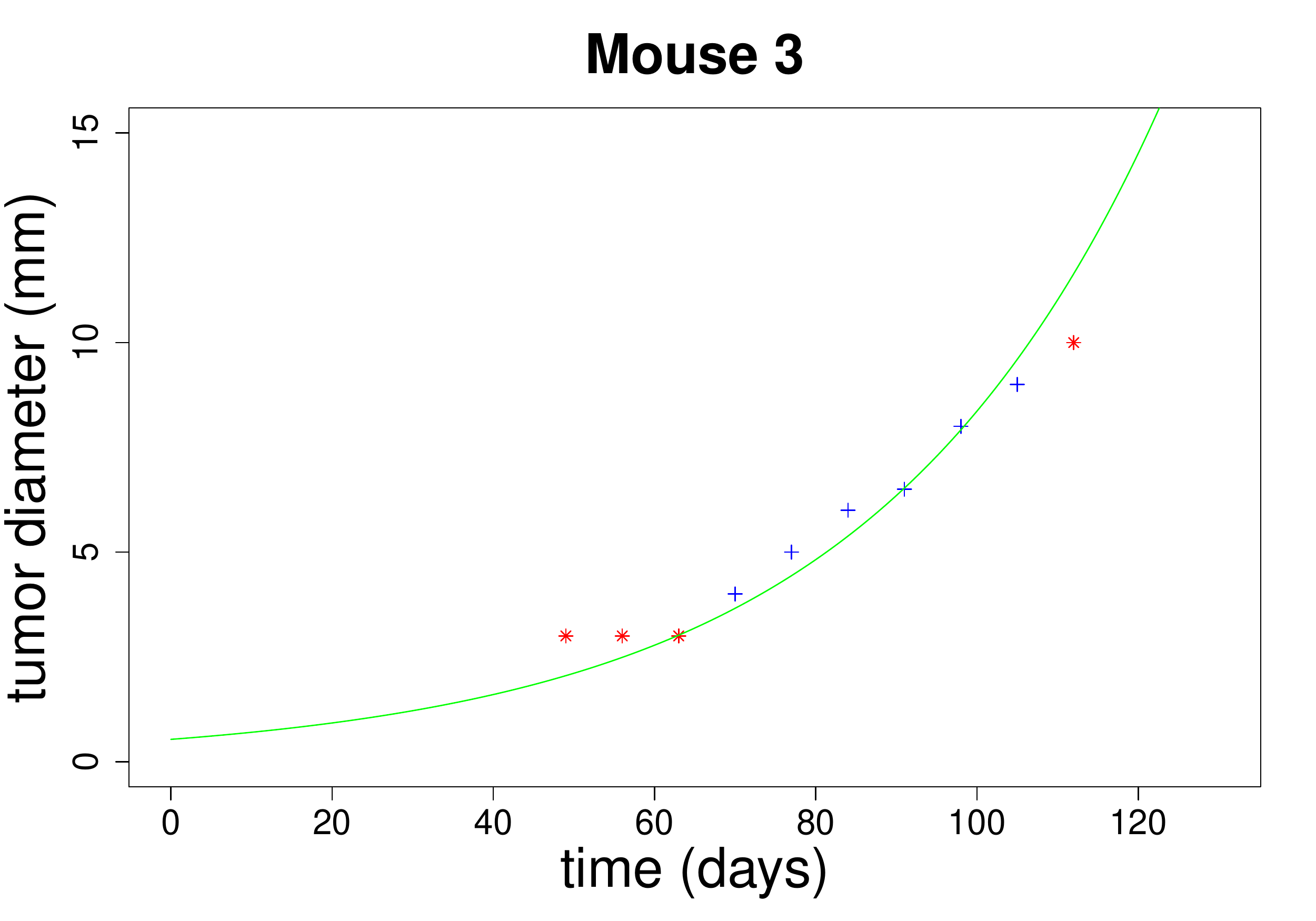}\hfill
    \includegraphics[width=.22\textwidth]{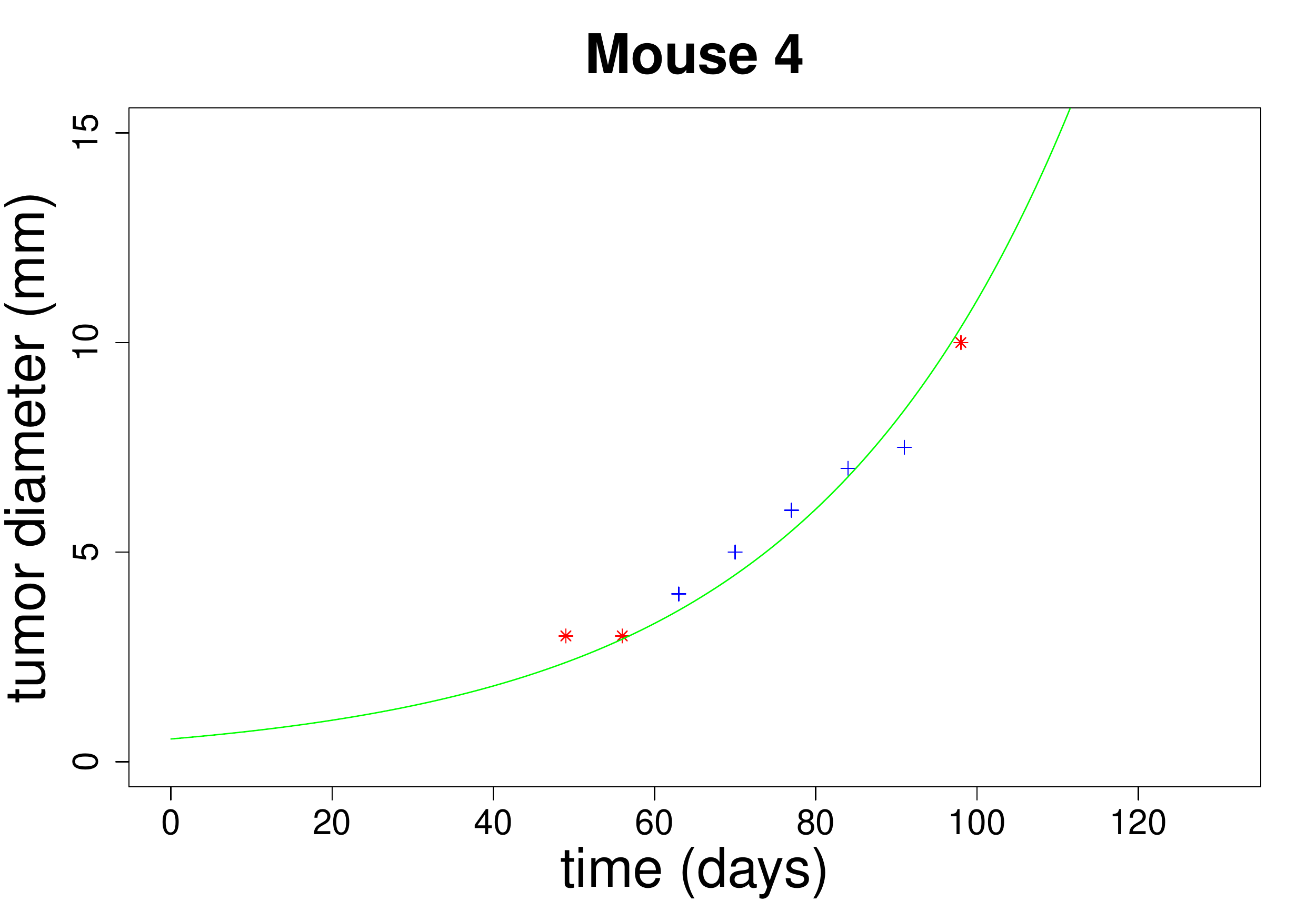}
    \\[\smallskipamount]
    \includegraphics[width=.22\textwidth]{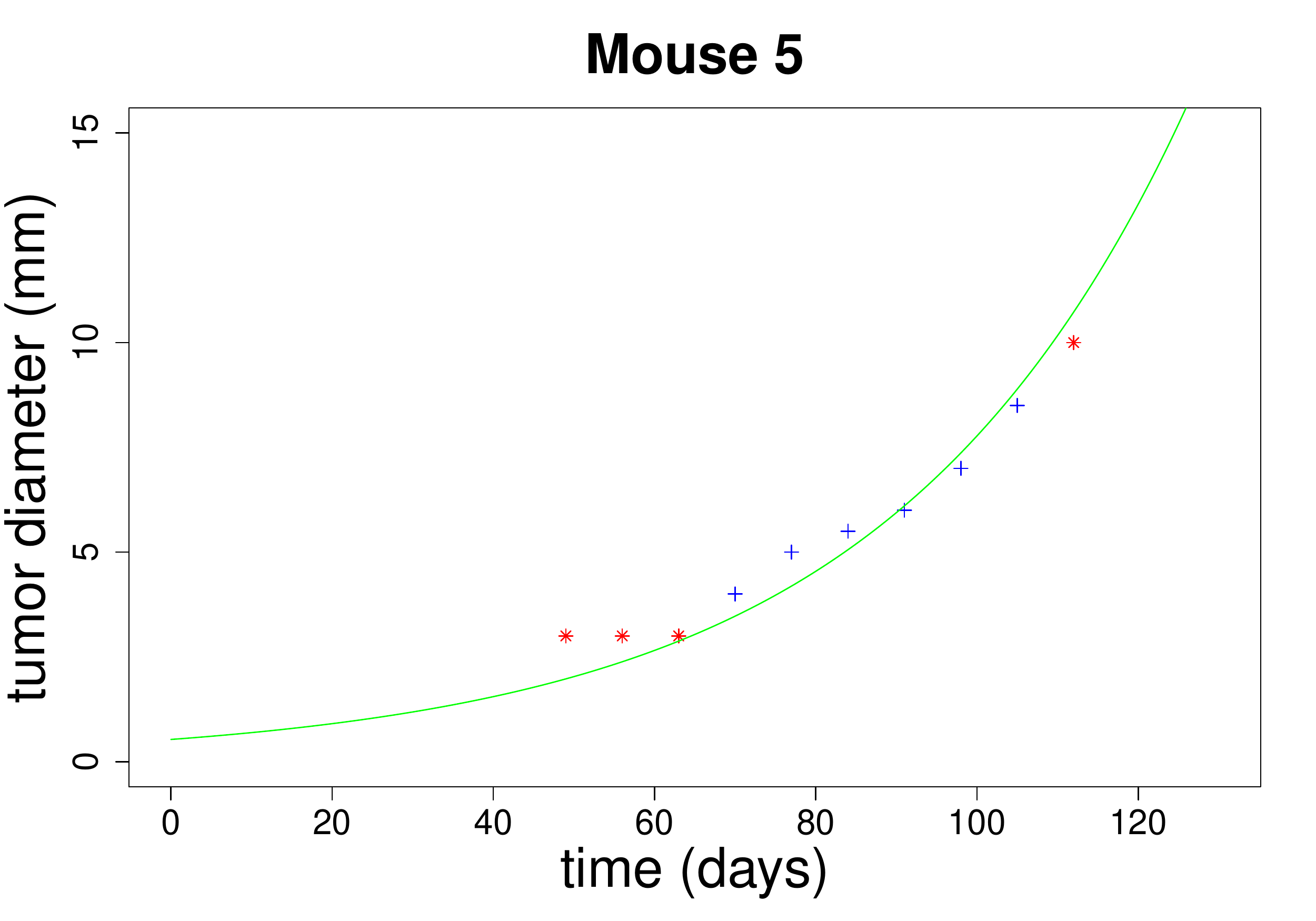}\hfill
    \includegraphics[width=.22\textwidth]{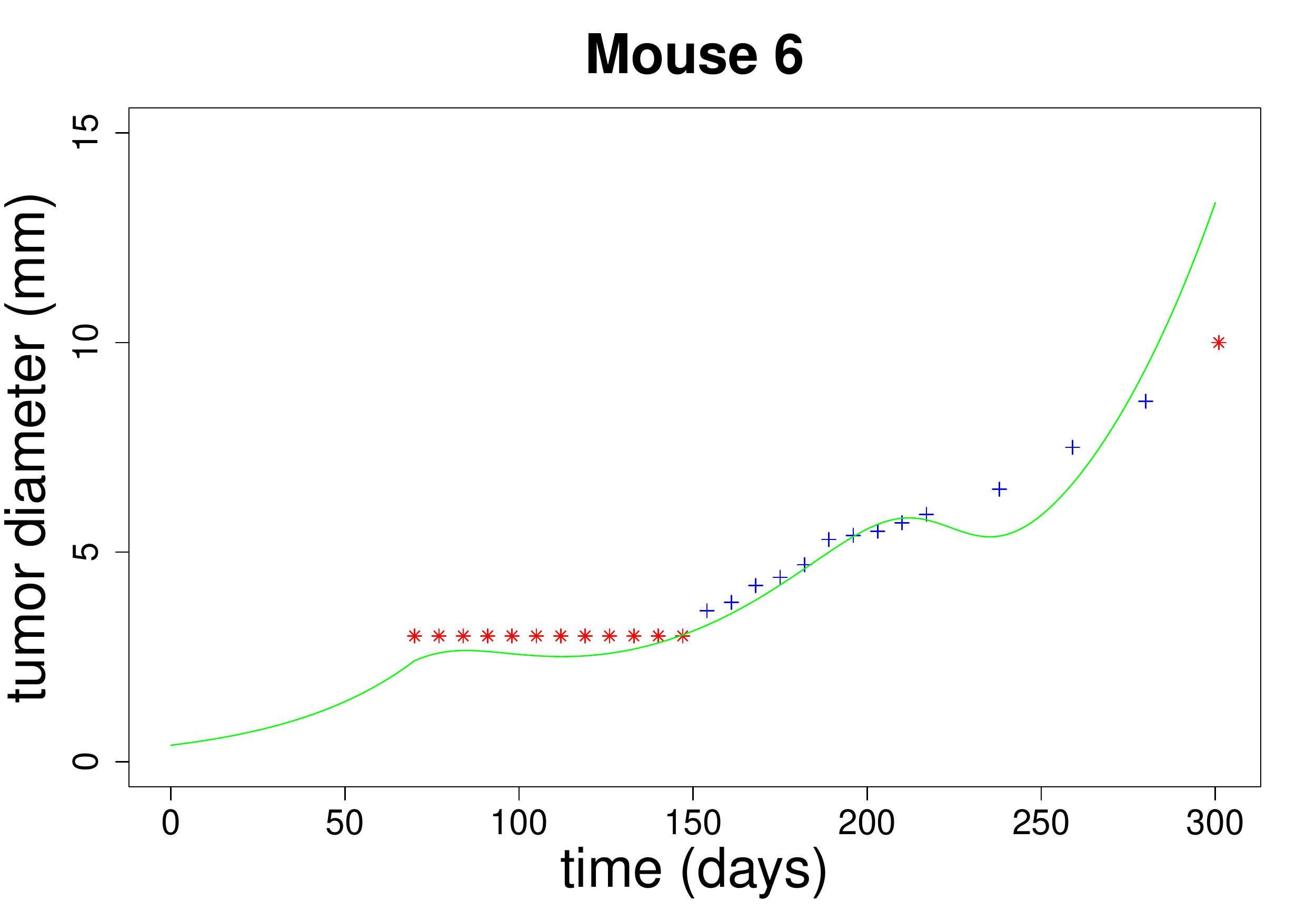}\hfill
    \includegraphics[width=.22\textwidth]{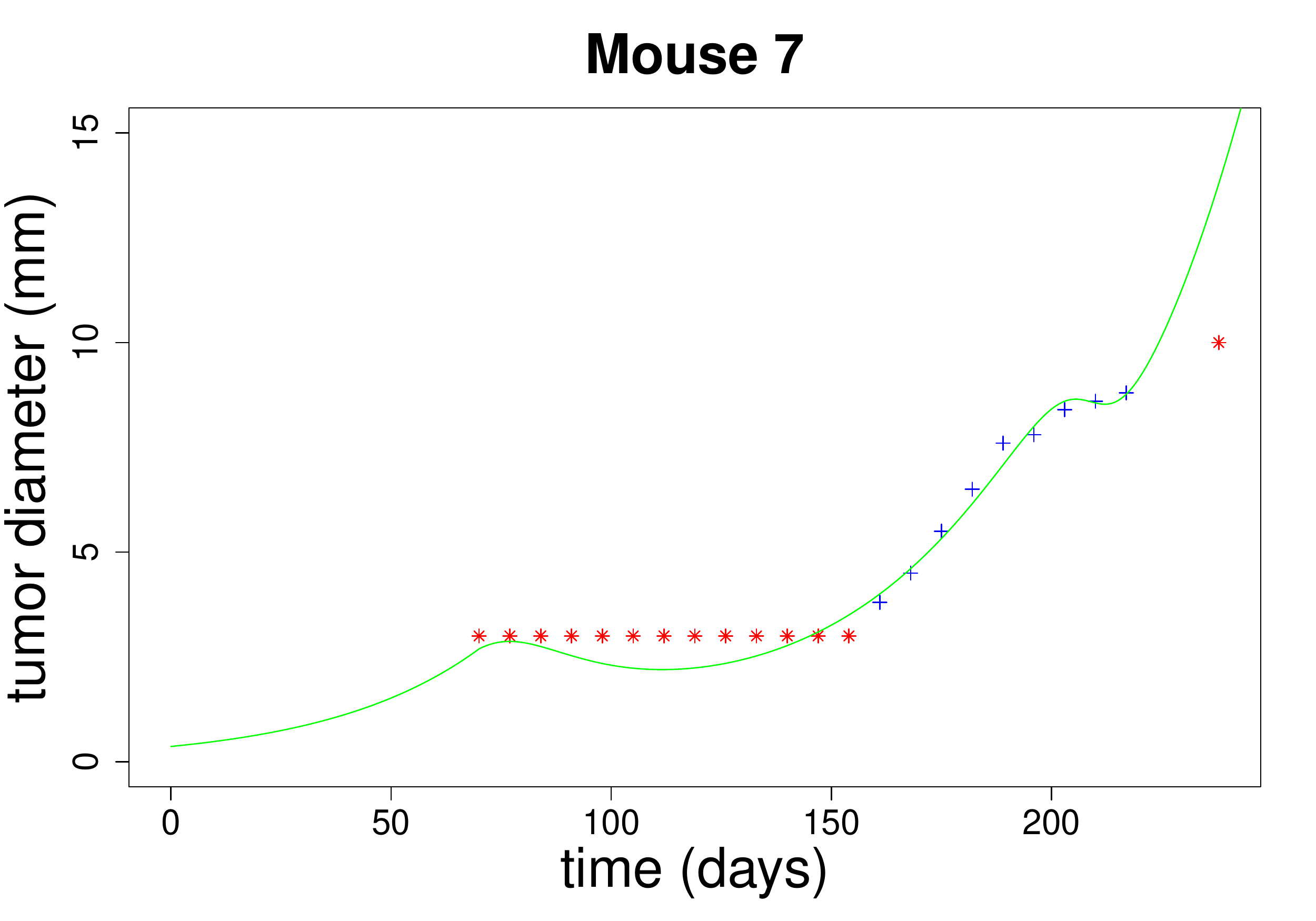}\hfill
    \includegraphics[width=.22\textwidth]{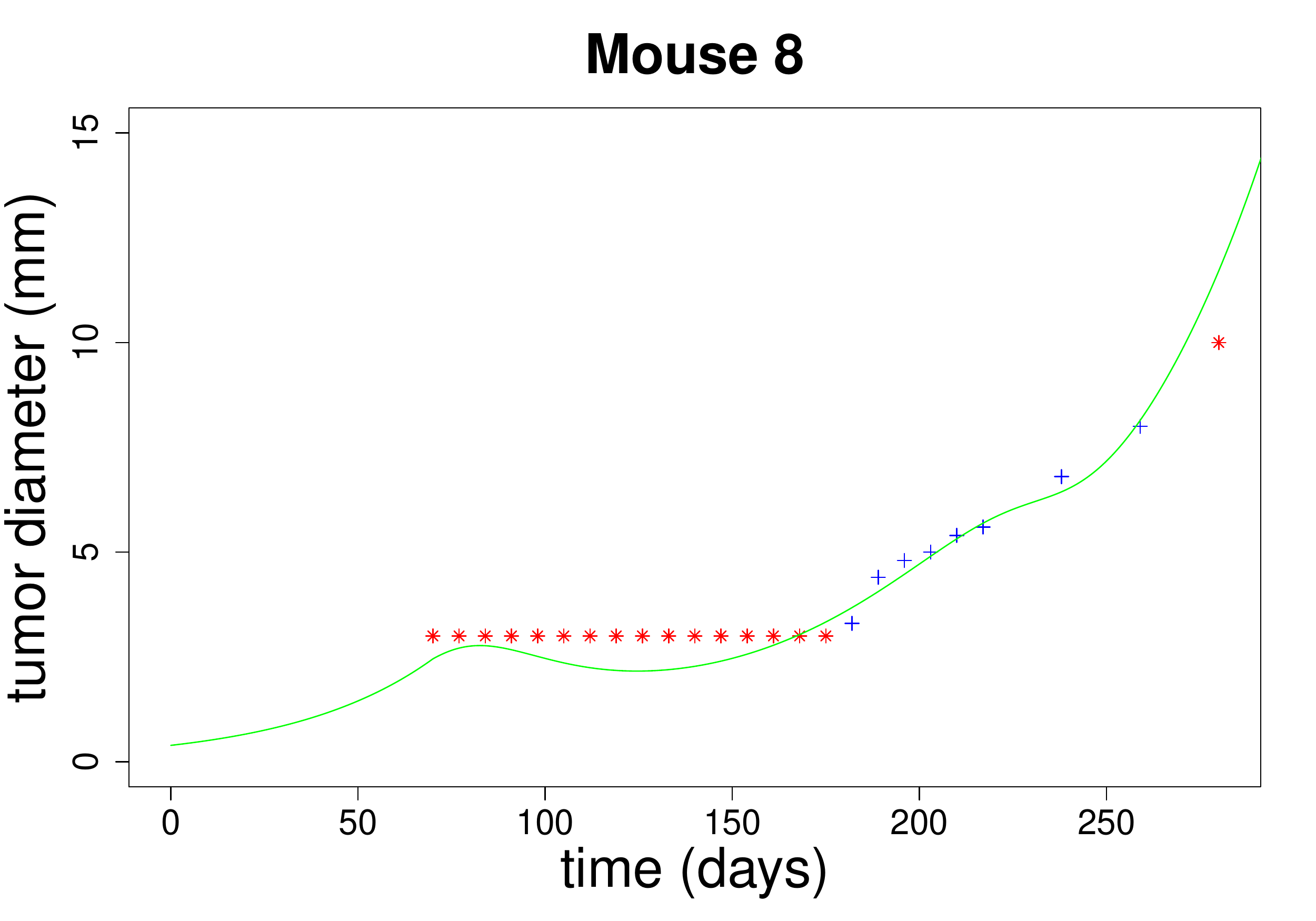}
    \\[\smallskipamount]
    \includegraphics[width=.22\textwidth]{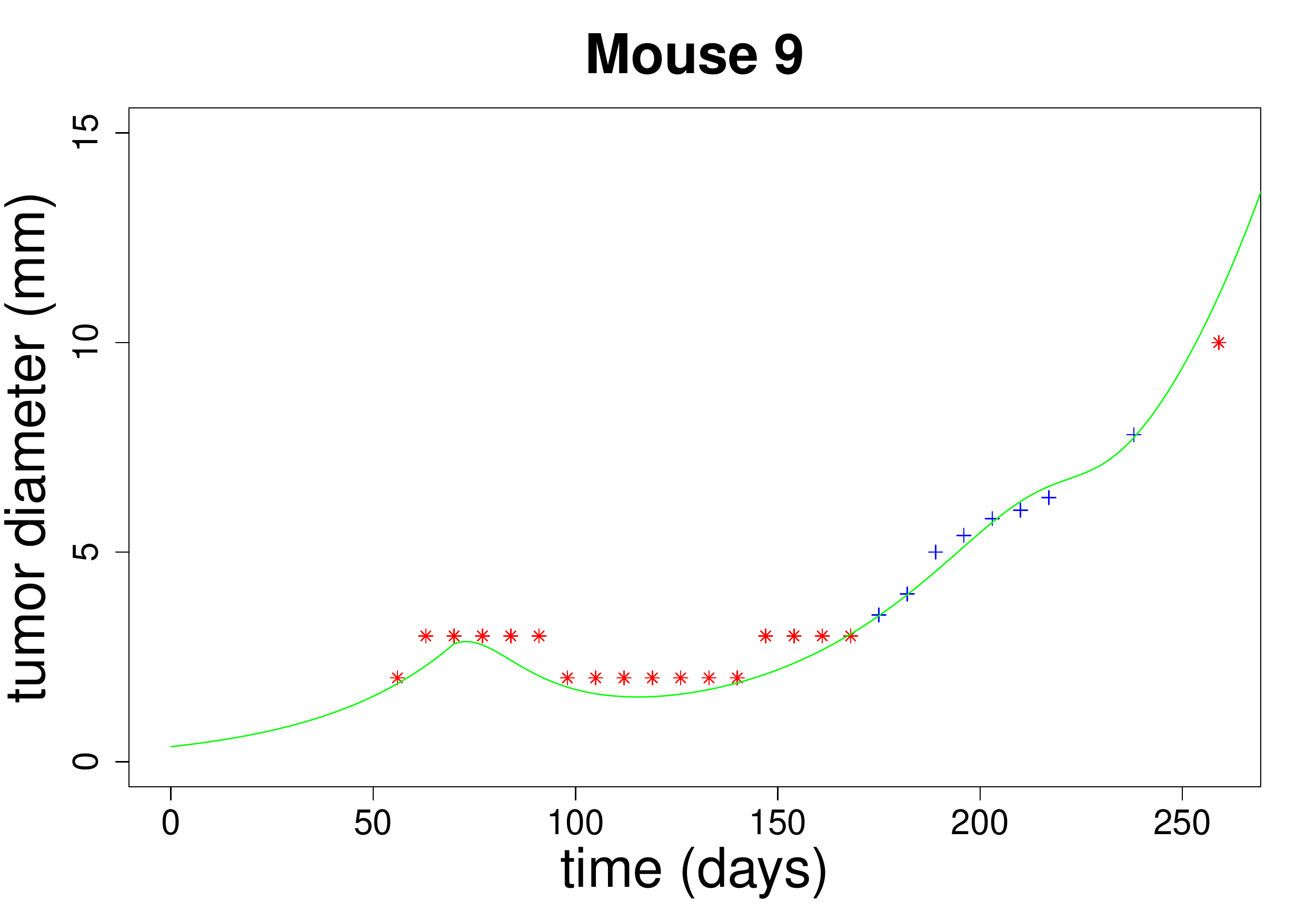}\hfill
    \includegraphics[width=.22\textwidth]{Images/LRT/Final/ind_fits_10.pdf}\hfill
    \includegraphics[width=.22\textwidth]{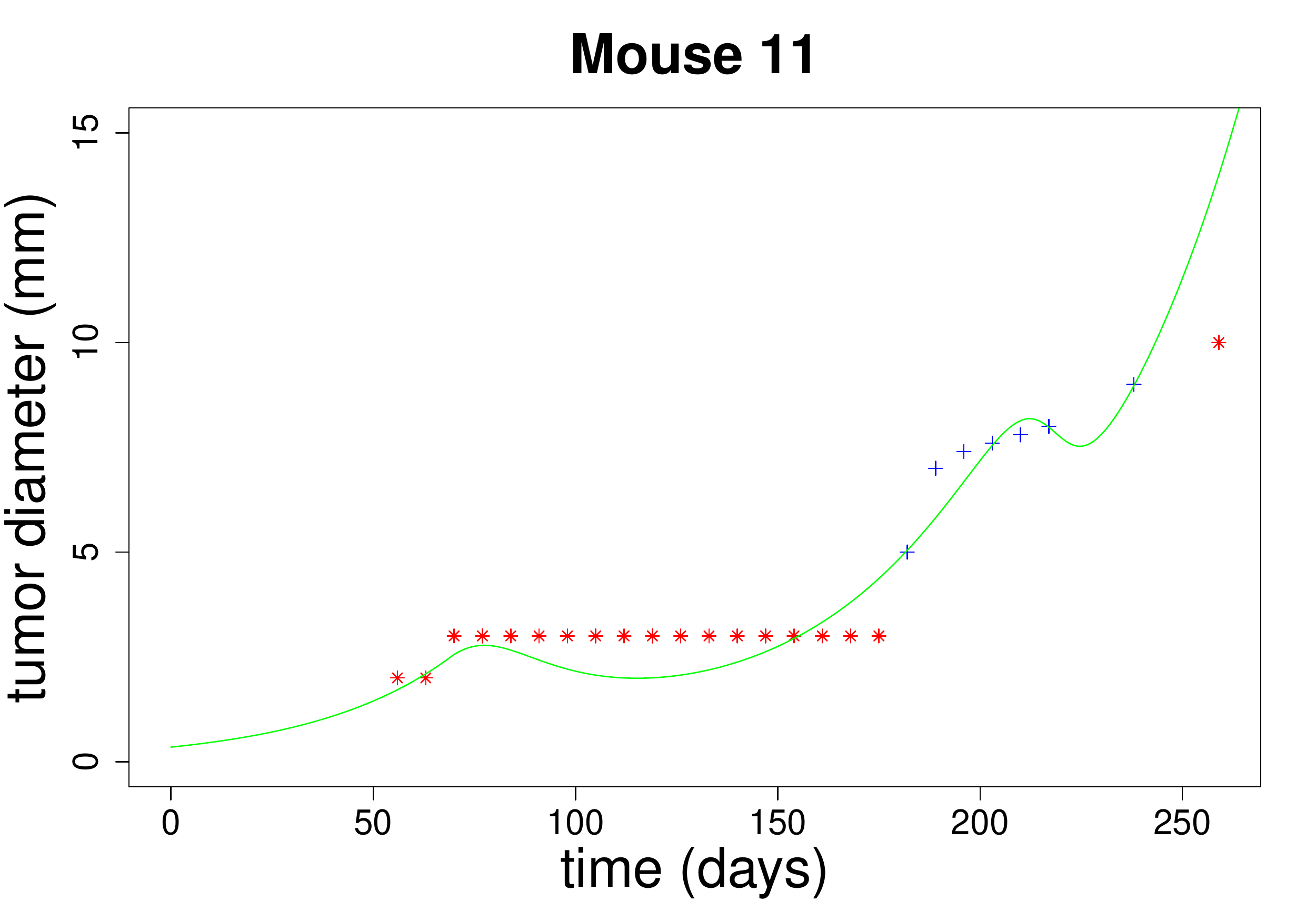}\hfill
    \includegraphics[width=.22\textwidth]{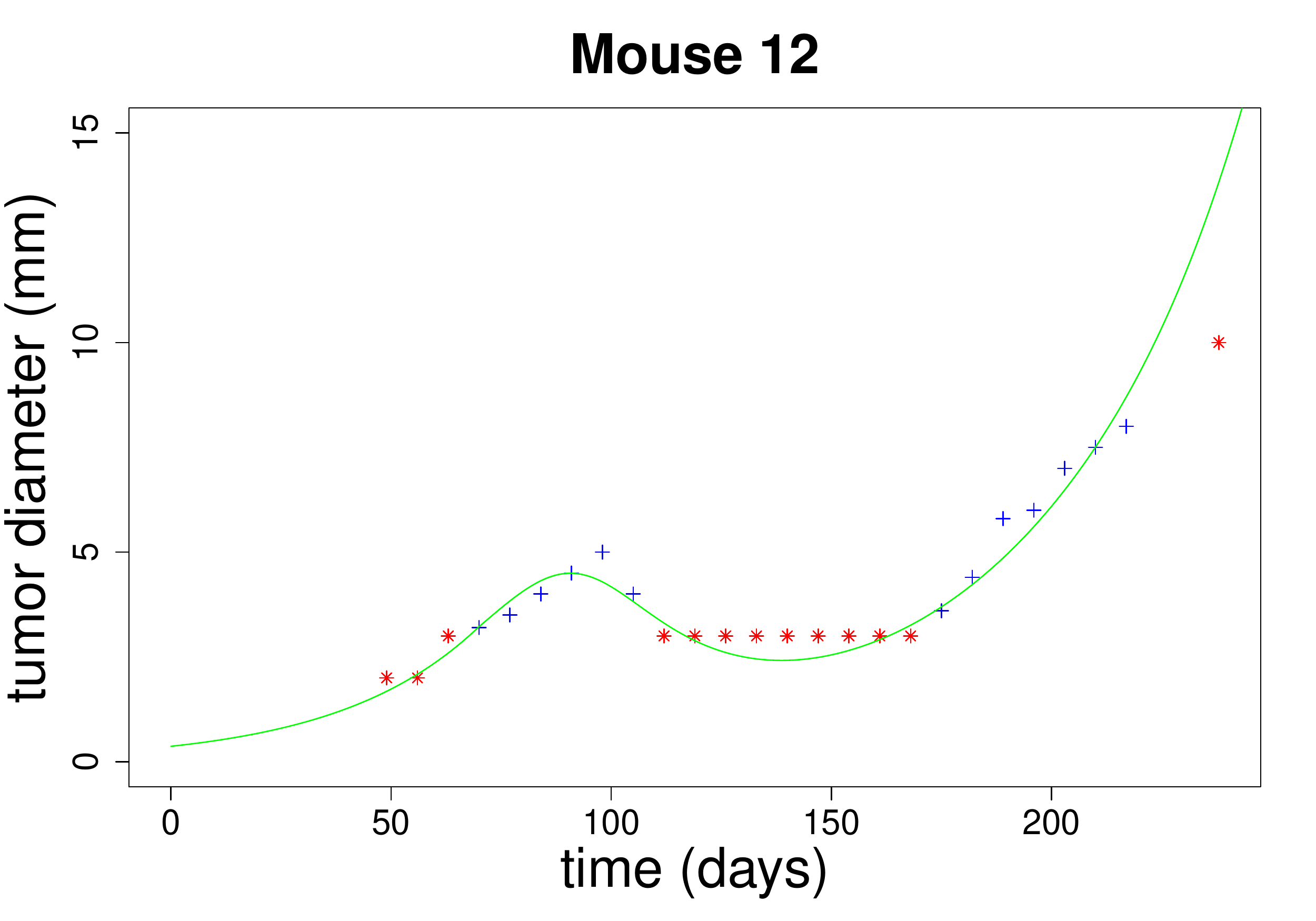}
    \\[\smallskipamount]
    \includegraphics[width=.22\textwidth]{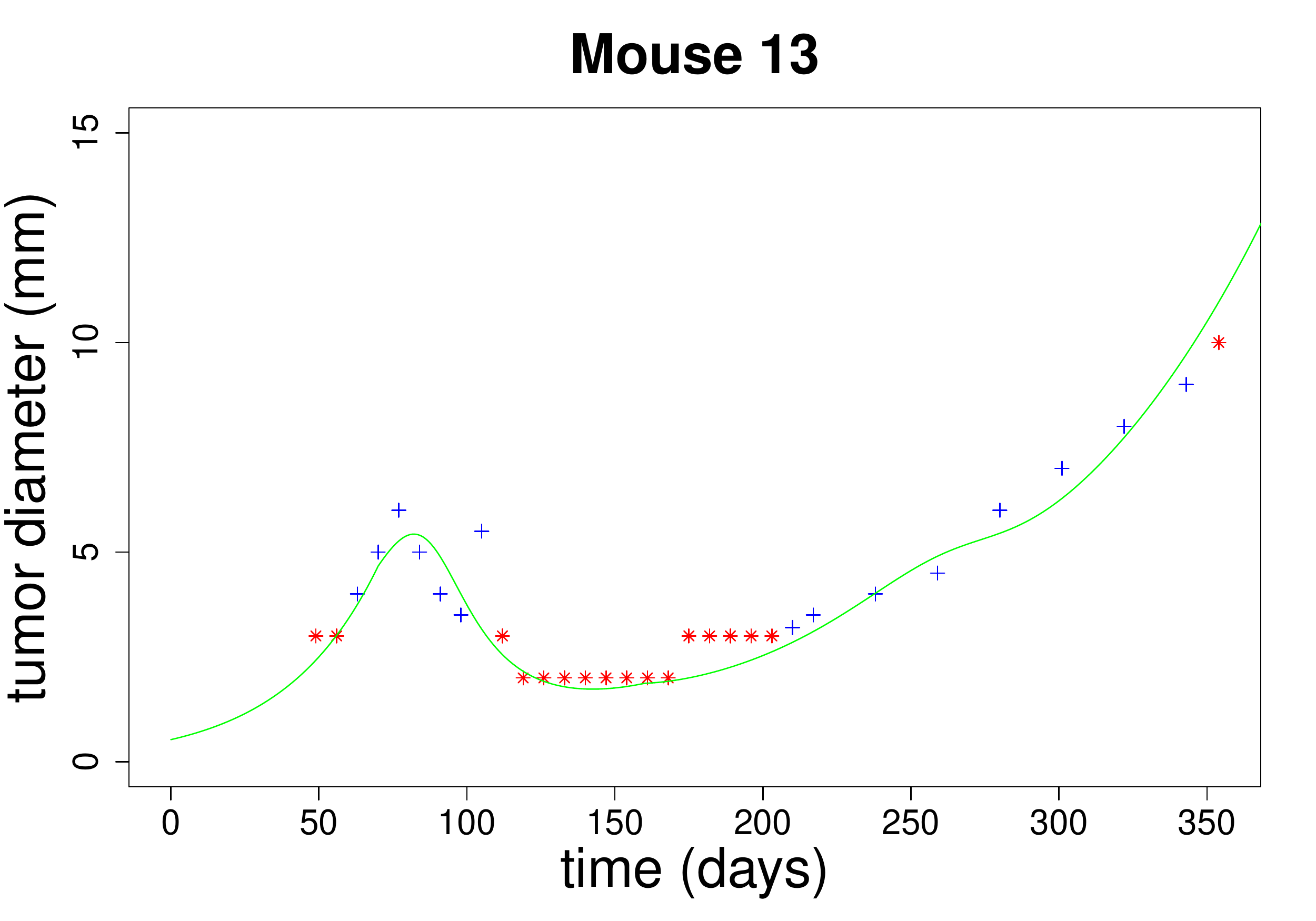}\hfill
    \includegraphics[width=.22\textwidth]{Images/LRT/Final/ind_fits_14.pdf}\hfill
    \includegraphics[width=.22\textwidth]{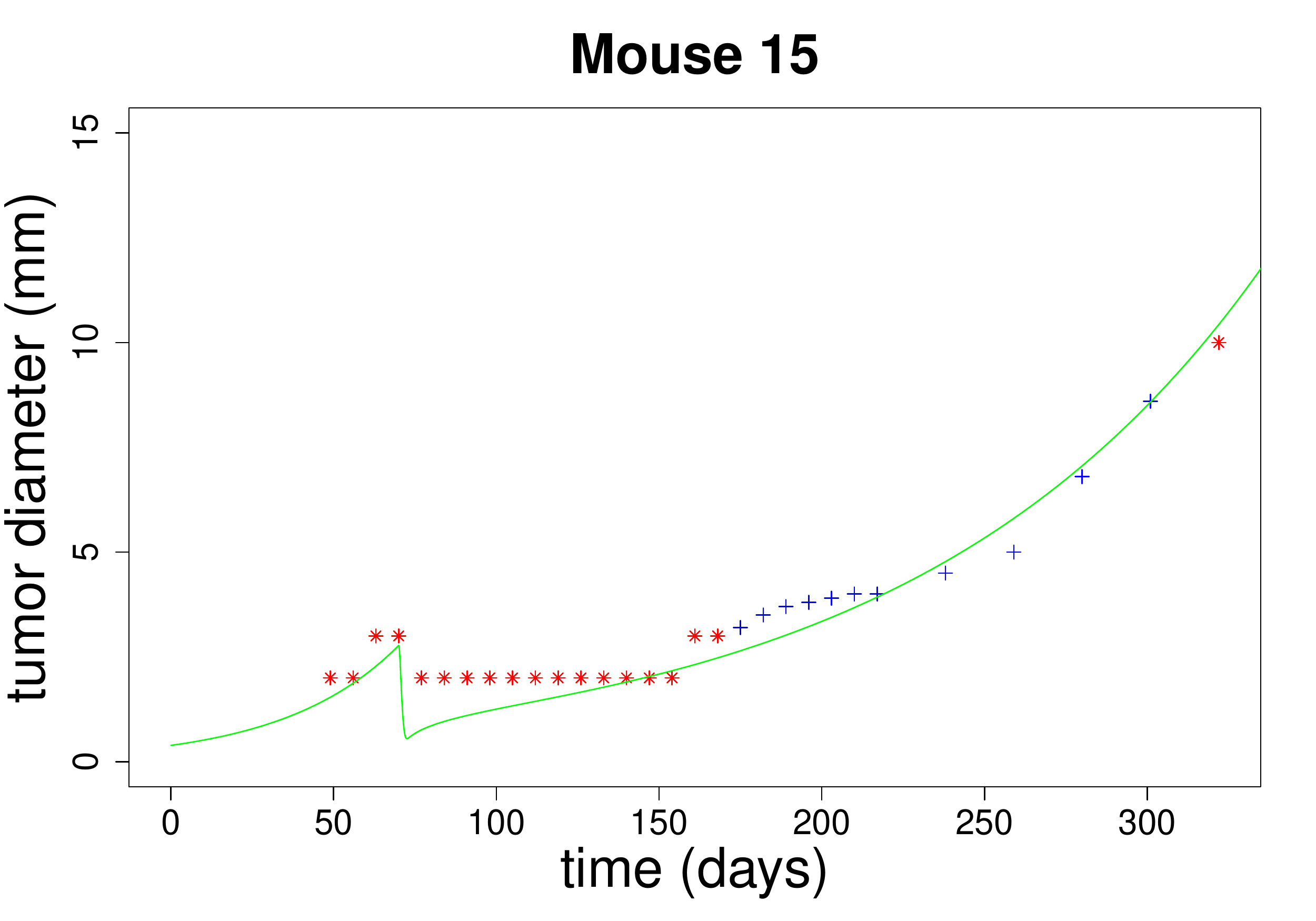}\hfill
    \includegraphics[width=.22\textwidth]{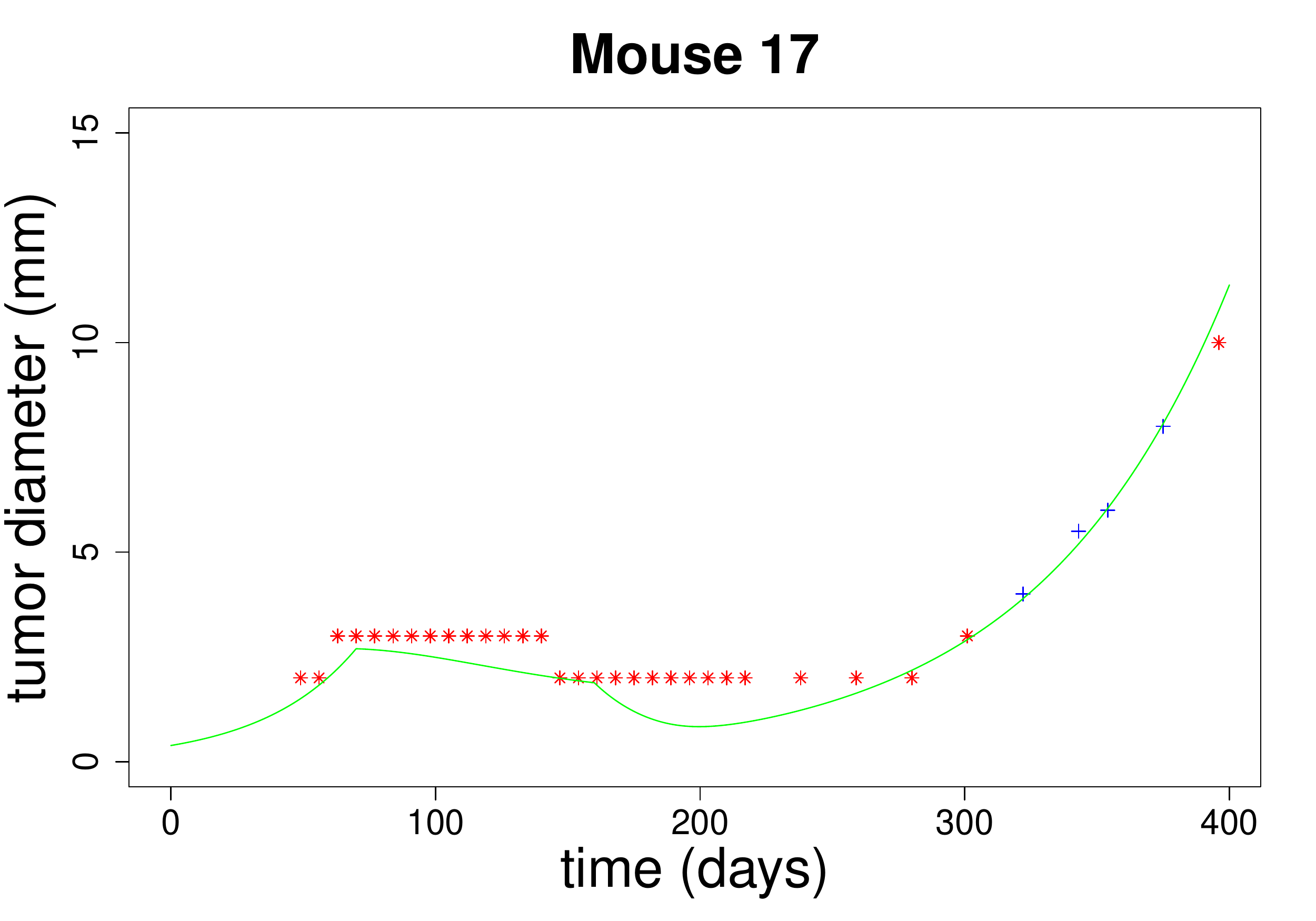}
        \\[\smallskipamount]
    \includegraphics[width=.22\textwidth]{Images/LRT/Final/ind_fits_18.pdf}\hfill
    \includegraphics[width=.22\textwidth]{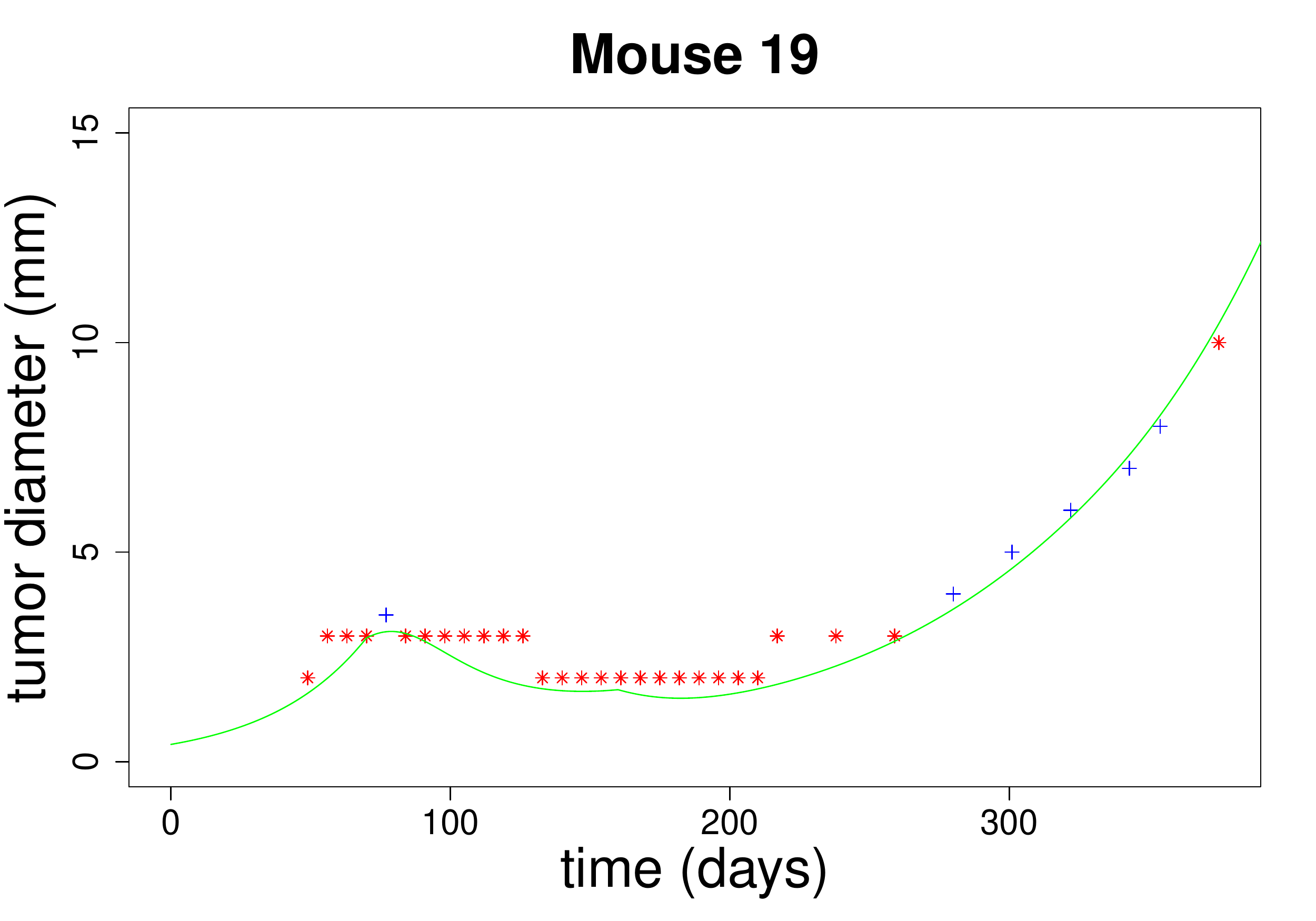}\hfill
    \includegraphics[width=.22\textwidth]{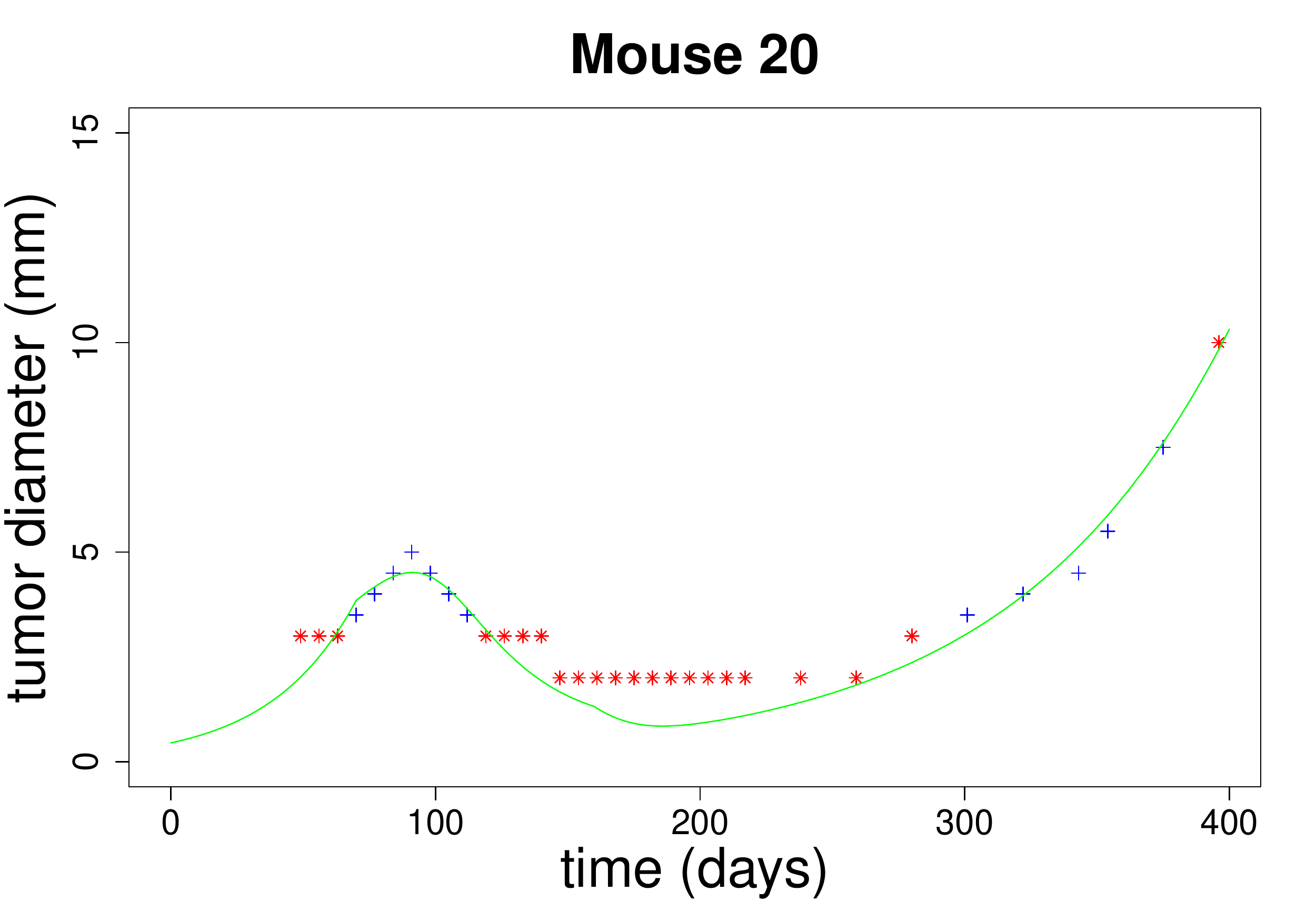}\hfill
    \includegraphics[width=.22\textwidth]{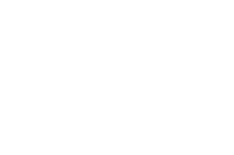}
    
  \caption{\scriptsize{Individual fits for the \textbf{final model}. Blue + marks: no censored observations,   red asterisk: censored observations,   green line: fitted tumor size dynamics. From left to right, top to bottom:  mice 1 to 5   in   CTRL group, mice 6 to 12   in   ACT group, and mice from 13 (last two rows)   in   ACT+Re group.}}
    \label{fits_all_inds_final}
\end{figure}


\begin{figure}
\begin{center}
  \includegraphics[height =0.45\linewidth , width=0.5\linewidth]{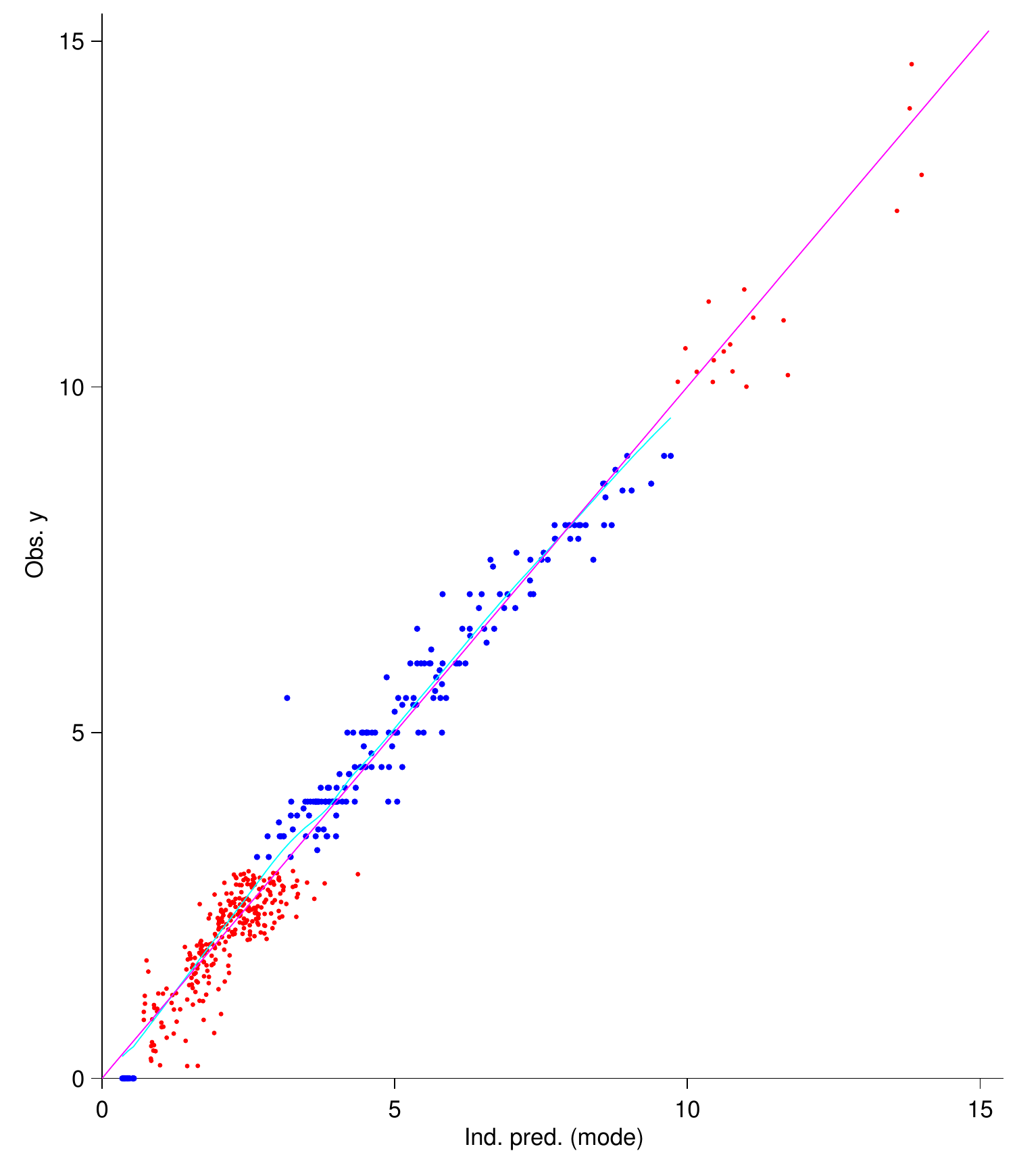}
  \caption{\scriptsize{Inidividual predicted values versus observed values  of the \textbf{final model}. Blue points: no censored observations,   red asterisk:   censored observations.}}
\label{pred_vs_obs_final}
  \end{center}
\end{figure}

\begin{figure}[H]
\begin{center}
  \includegraphics[height =0.5\linewidth, width=0.9\linewidth]{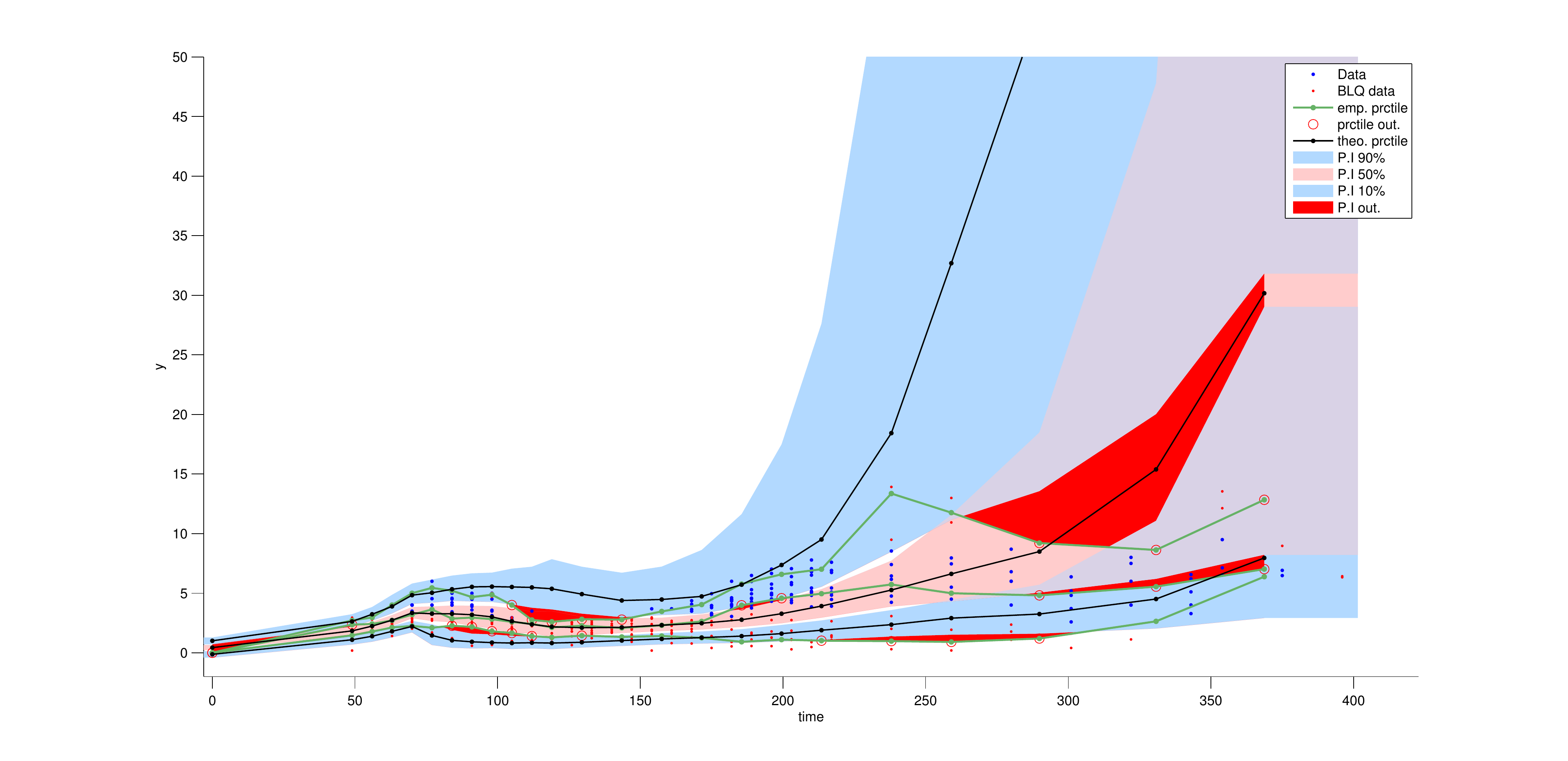}
  \caption{\scriptsize{Prediction-corrected Visual Predictive Checks (pcVPC) for the \textbf{final model}.}}
  \label{pcVPC_final}
  \end{center}
\end{figure}
\begin{figure} 
		\centering\includegraphics[width=0.46\textwidth]{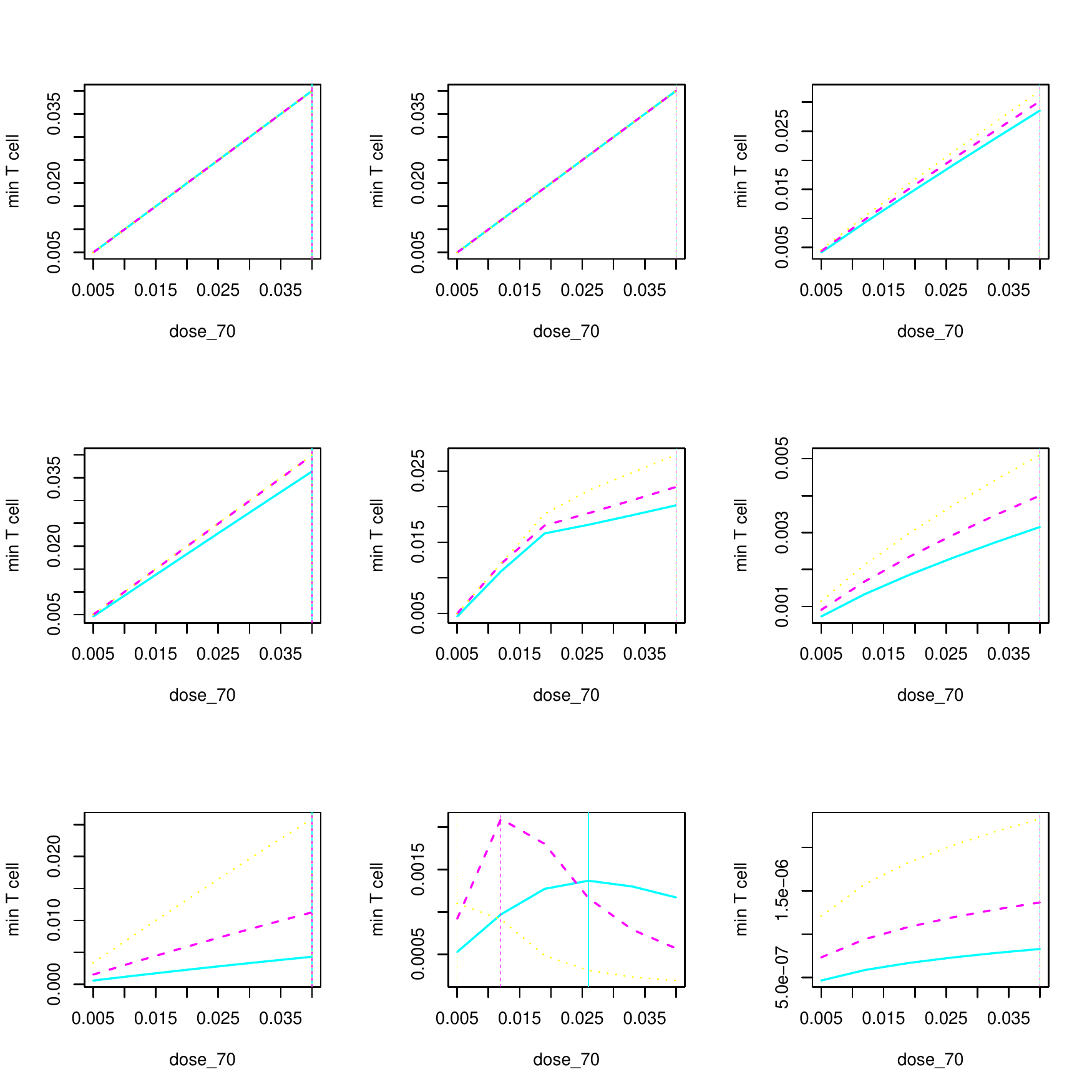}
	\caption{\scriptsize{T cell minimum according to treatment dose $d_{70}$ in ACT group for $d_{MT} = \text{q}_{5}(t_{T_i})$. 
	Top-figures: $d_T = \text{q}_{5}(d_{T_i})$ ; {\color{black}Middle-figures}: $d_T = \text{q}_{50}(d_{T_i})$ ; Bottom-figures: $d_T = \text{q}_{95}(d_{T_i})$. Left-figures: $l_{A}^{\text{prod}} = \text{q}_{5}(l_{A_i}^{\text{prod}})$  ; {\color{black}Middle-figures}:  $l_{A}^{\text{prod}} = \text{q}_{50}(l_{A_i}^{\text{prod}})$ ; Right-figures:  $l_{A}^{\text{prod}} = \text{q}_{95}(l_{A_i}^{\text{prod}})$. In each sub-figure: $n_{M_{0}} = \text{q}_{5}(n_{M_{0_i}})$ (solid curve in cyan);  $n_{M_{0}} = \text{q}_{50}(n_{M_{0_i}})$ (dashed curve in purple);  $n_{M_{0}} = \text{q}_{95}(n_{M_{0_i}})$ (dotted curve in yellow). The vertical solid, dashed and dotted lines indicate the optimal dose $d_{70}^{\text{opt}_{*}}$ for a given $n_{M_{0}}$.}}	
	\label{minTc_vs_dose70_ACT_tT_5}
\end{figure}

\begin{figure} 
		\centering\includegraphics[width=0.46\textwidth]{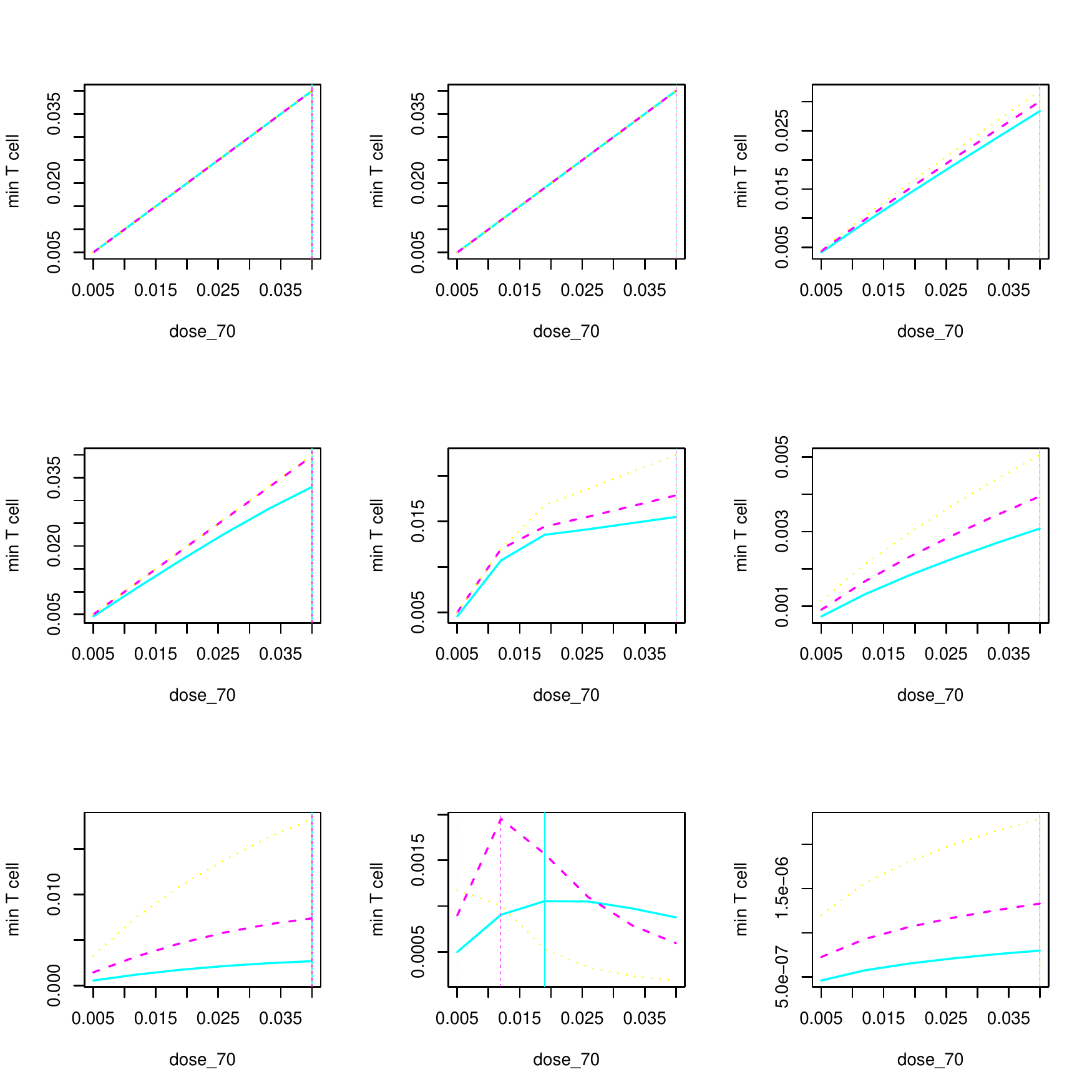}
	\caption{\scriptsize{T cell minimum according to treatment dose $d_{70}$ in ACT group for $d_{MT} = \text{q}_{50}(t_{T_i})$. 
	Top-figures: $d_T = \text{q}_{5}(d_{T_i})$ ; {\color{black}Middle-figures}: $d_T = \text{q}_{50}(d_{T_i})$ ; Bottom-figures: $d_T = \text{q}_{95}(d_{T_i})$. Left-figures: $l_{A}^{\text{prod}} = \text{q}_{5}(l_{A_i}^{\text{prod}})$  ; {\color{black}Middle-figures}:  $l_{A}^{\text{prod}} = \text{q}_{50}(l_{A_i}^{\text{prod}})$ ; Right-figures:  $l_{A}^{\text{prod}} = \text{q}_{95}(l_{A_i}^{\text{prod}})$.In each sub-figure: $n_{M_{0}} = \text{q}_{5}(n_{M_{0_i}})$ (solid curve in cyan);  $n_{M_{0}} = \text{q}_{50}(n_{M_{0_i}})$ (dashed curve in purple);  $n_{M_{0}} = \text{q}_{95}(n_{M_{0_i}})$ (dotted curve in yellow). The vertical solid, dashed and dotted lines indicate the optimal dose $d_{70}^{\text{opt}_{*}}$ for a given $n_{M_{0}}$.}}
	\label{minTc_vs_dose70_ACT_tT_50_quantiles_5_50_95}
\end{figure}

\begin{figure} 
		\centering\includegraphics[width=0.46\textwidth]{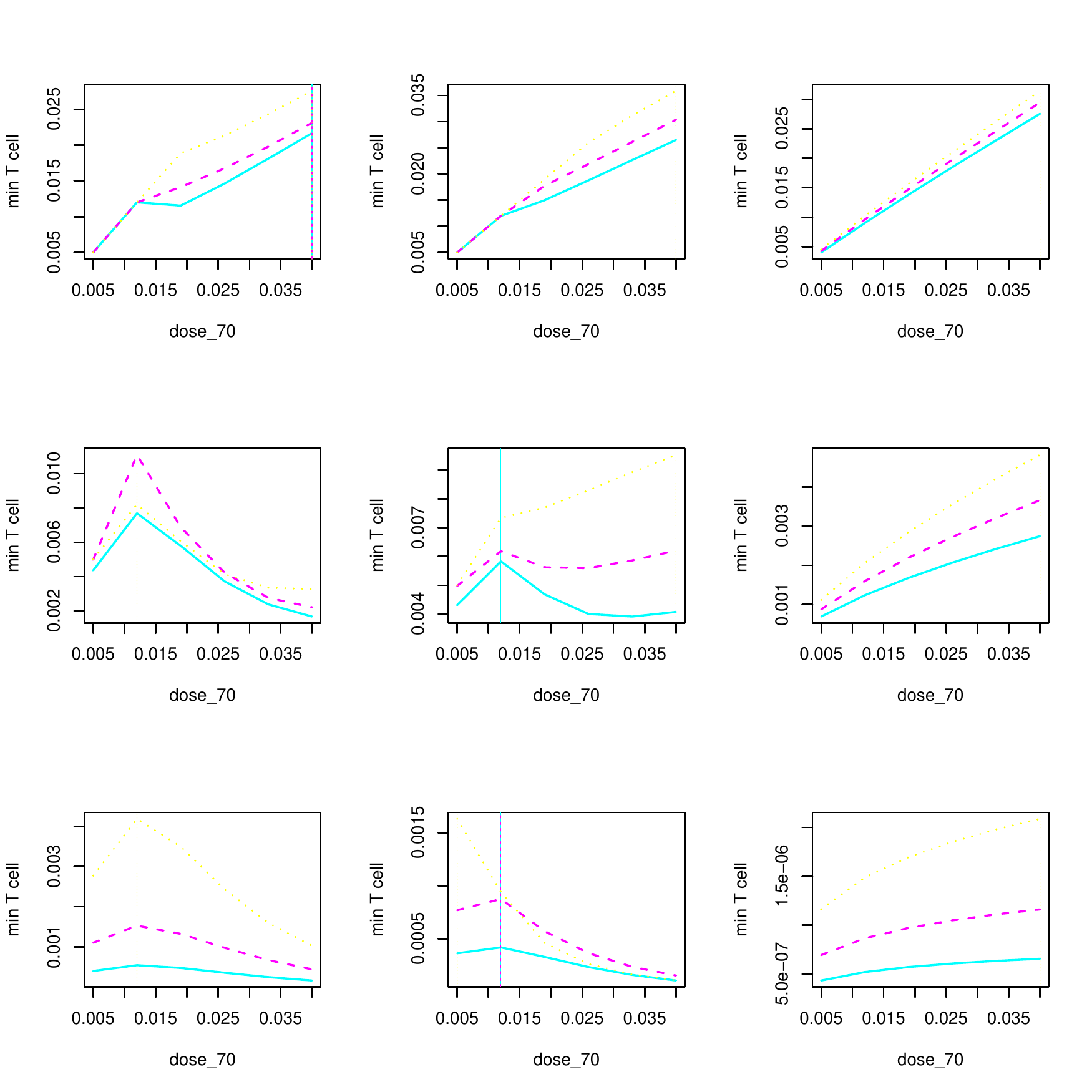}
	\caption{\scriptsize{T cell minimum according to treatment dose $d_{70}$ in ACT group for $d_{MT} = \text{q}_{95}(t_{T_i})$. 
	Top-figures: $d_T = \text{q}_{5}(d_{T_i})$ ; {\color{black}Middle-figures}: $d_T = \text{q}_{50}(d_{T_i})$ ; Bottom-figures: $d_T = \text{q}_{95}(d_{T_i})$. Left-figures: $l_{A}^{\text{prod}} = \text{q}_{5}(l_{A_i}^{\text{prod}})$  ; {\color{black}Middle-figures}:  $l_{A}^{\text{prod}} = \text{q}_{50}(l_{A_i}^{\text{prod}})$ ; Right-figures:  $l_{A}^{\text{prod}} = \text{q}_{95}(l_{A_i}^{\text{prod}})$.In each sub-figure: $n_{M_{0}} = \text{q}_{5}(n_{M_{0_i}})$ (solid curve in cyan);  $n_{M_{0}} = \text{q}_{50}(n_{M_{0_i}})$ (dashed curve in purple);  $n_{M_{0}} = \text{q}_{95}(n_{M_{0_i}})$ (dotted curve in yellow). The vertical solid, dashed and dotted lines indicate the optimal dose $d_{70}^{\text{opt}_{*}}$ for a given $n_{M_{0}}$.}}	
	\label{minTc_vs_dose70_ACT_tT_95}
\end{figure}

\begin{figure} 
		\centering\includegraphics[width=0.44\textwidth]{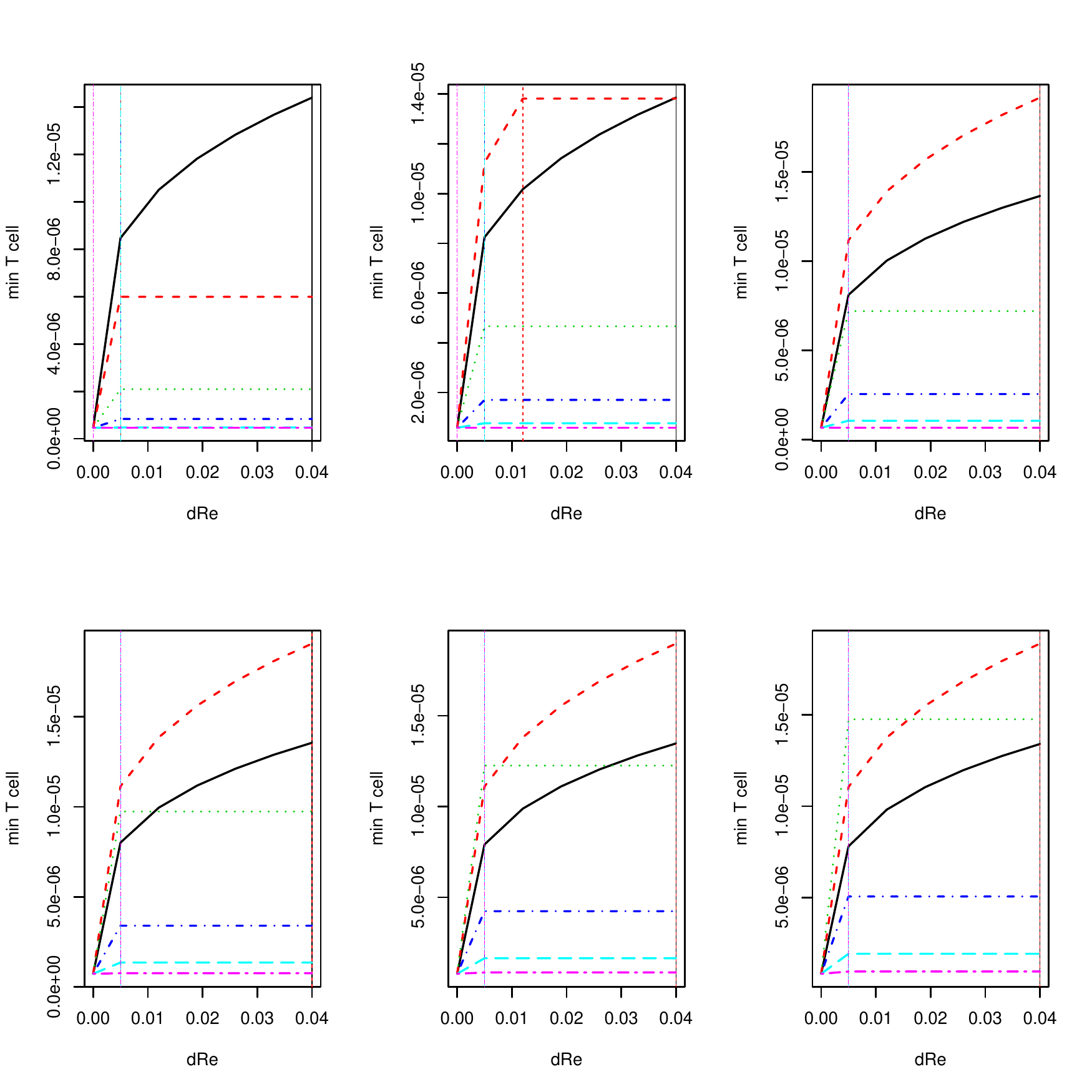}
	\caption{\scriptsize{T cell minimum according to treatment parameters in ACTRE group for $d_{MT} = \text{q}_{5}(t_{T_i})$; $d_T = \text{q}_{95}(d_{T_i})$; $l_{A}^{\text{prod}} = \text{q}_{95}(l_{A_i}^{\text{prod}})$; $n_{M_{0}} = \text{q}_{5}(n_{M_{0_i}})$. From top to bottom and from left to right: $d_{70} =  0.005, \ 0.012, \ 0.019, \ 0.026, \ 0.033, \ 0.040$. 
	In each subfigure: curves for   discrete retreatment times $t_{\text{Re}}=130$  (black solid), 142 (red dashed), 154 (green dotted), 166  (blue dotdashed), 178 (cyan longdashed), 190 (purple twodashed). Vertical lines: optimal $d_{\text{Re}}^{\text{opt}}$.   $d_{70}^{\text{opt}_{**}} = 0.019, d_{\text{Re}}^{\text{opt}} = 0.04, t_{\text{Re}}^{\text{opt}} = 142$.}}
	\label{minTc_vs_dose70_tRe_dRe_ACTRE_quantiles_5_50_95_interessant2}
\end{figure}

\begin{figure} 
		\centering\includegraphics[width=0.44\textwidth]{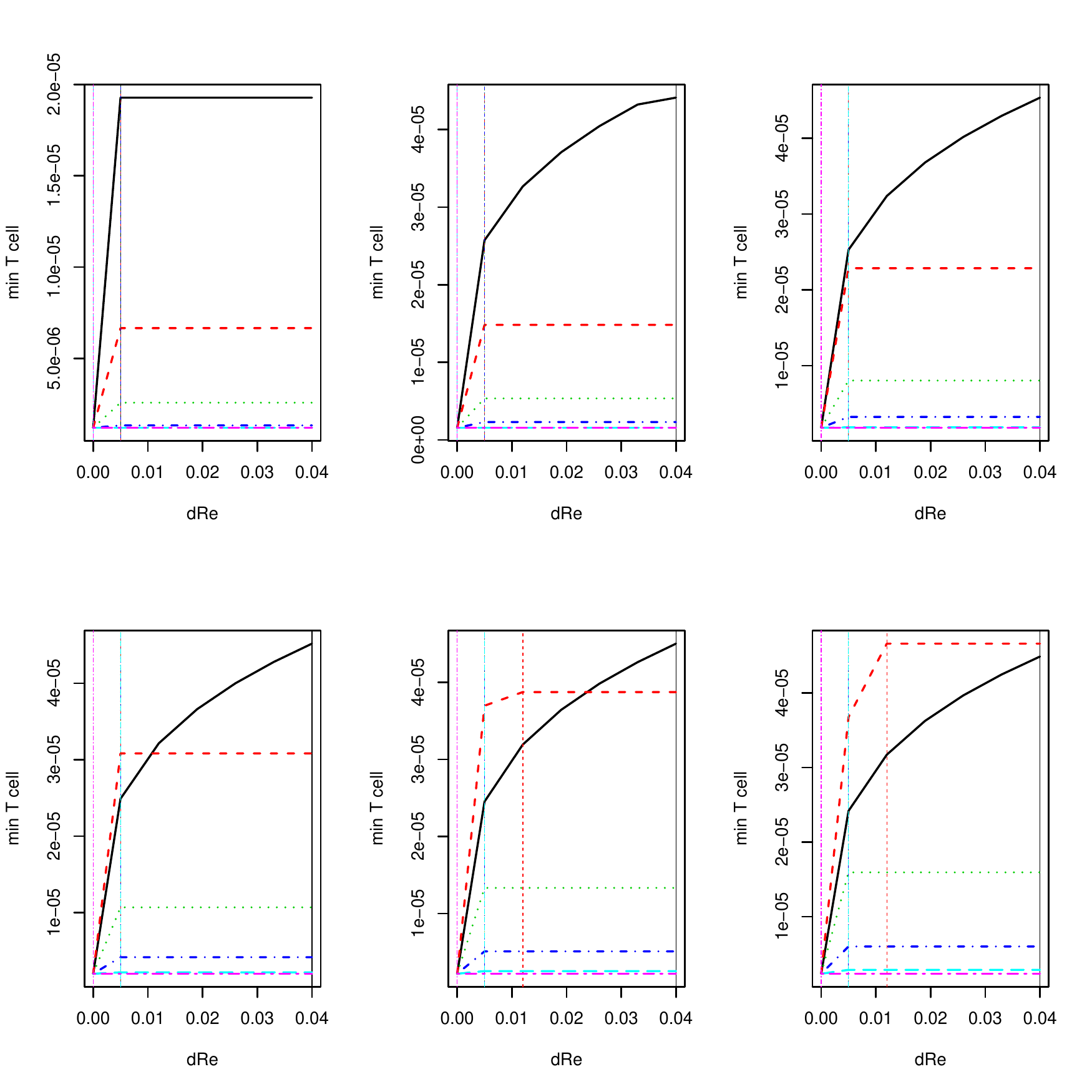}
	\caption{\scriptsize{T cell minimum according to treatment parameters in ACTRE group for $d_{MT} = \text{q}_{5}(t_{T_i})$; $d_T = \text{q}_{95}(d_{T_i})$; $l_{A}^{\text{prod}} = \text{q}_{95}(l_{A_i}^{\text{prod}})$; $n_{M_{0}} = \text{q}_{95}(n_{M_{0_i}})$. From top to bottom and from left to right: $d_{70} =  0.005, \ 0.012, \ 0.019, \ 0.026, \ 0.033, \ 0.040$. In each subfigure: curves for   discrete retreatment times $t_{\text{Re}}=130$  (black solid), 142 (red dashed), 154 (green dotted), 166  (blue dotdashed), 178 (cyan longdashed), 190 (purple twodashed). Vertical lines: optimal $d_{\text{Re}}^{\text{opt}}$.   $d_{70}^{\text{opt}_{**}} = 0.04, d_{\text{Re}}^{\text{opt}} = 0.012, t_{\text{Re}}^{\text{opt}} = 142$.}}
	\label{minTc_vs_dose70_tRe_dRe_ACTRE_quantiles_5_50_95_interessant}
\end{figure}

 \begin{figure} 
		\centering\includegraphics[width=0.45\textwidth]{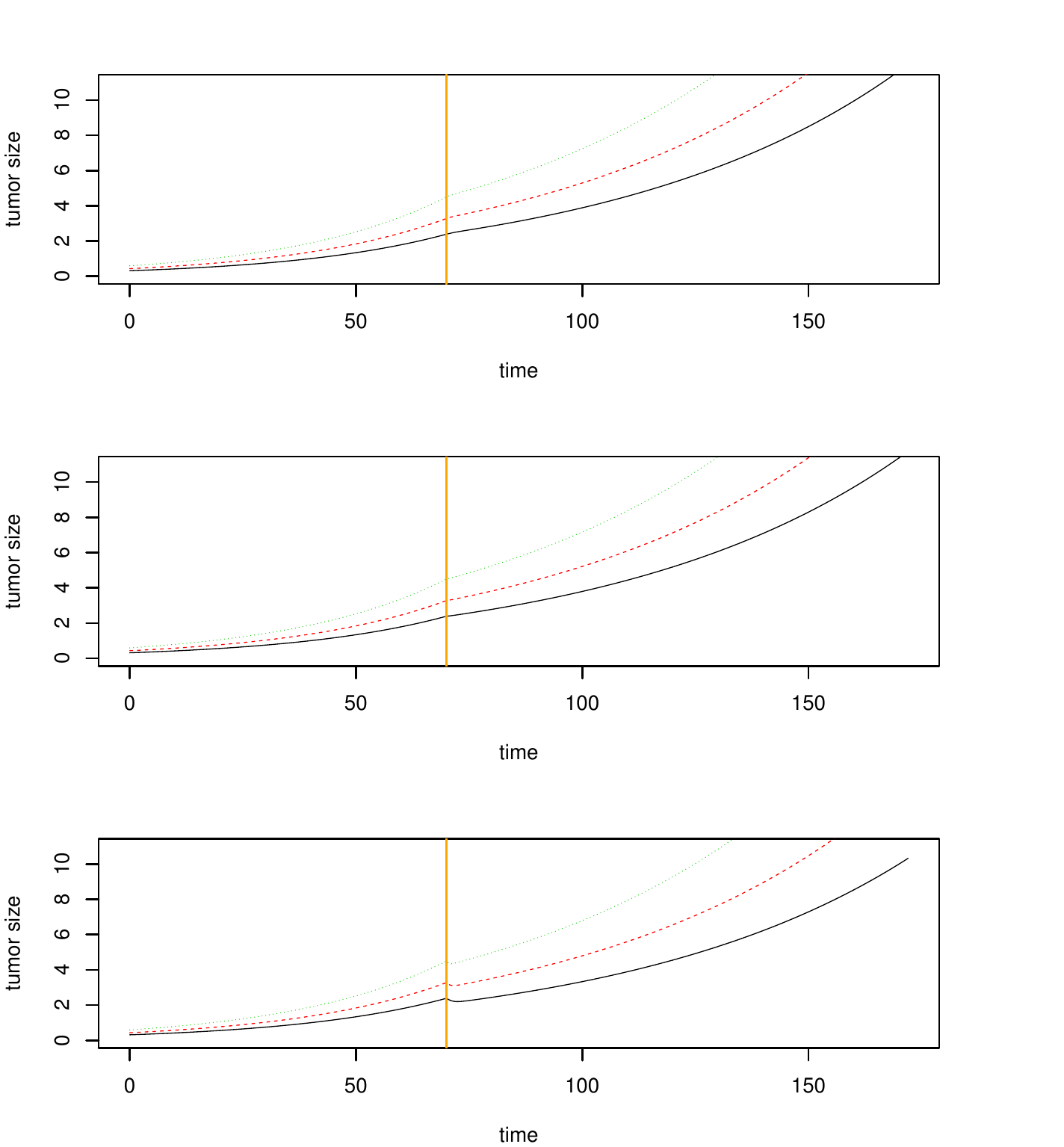}
	\caption{\scriptsize{Tumor size along time using quantiles $\text{q}_{\alpha}(\psi_i)$, $\alpha \in \{5, 50, 95\}$, for $d_{TM}$, $d_{T}$, $l_{A}^{\text{prod}}$, $n_{M_{0}}$ with  $d_{70} = d_{\text{max}}$, $d_T = \text{q}_{95}(d_{T})$, $l_{A}^{\text{prod}} = \text{q}_{95}(l_{A}^{\text{prod}})$. Top: $d_{MT} = \text{q}_{5}(d_{MT})$; Middle: $d_{MT} = \text{q}_{50}(d_{MT})$; Bottom: $d_{MT} = \text{q}_{95}(d_{MT})$. For each subfigure,  $n_{M_0} = \text{q}_{5}(n_{M_{0_i}})$ in black solid line, $n_{M_0} = \text{q}_{50}(n_{M_{0_i}})$ in red dashed line, $n_{M_0} = \text{q}_{95}(n_{M_{0_i}})$ in green dotted line. Vertical orange line:   treatment start.}}
	\label{tumorSize_vs_time}
\end{figure} 

\begin{figure} 
		\centering\includegraphics[width=0.54\textwidth]{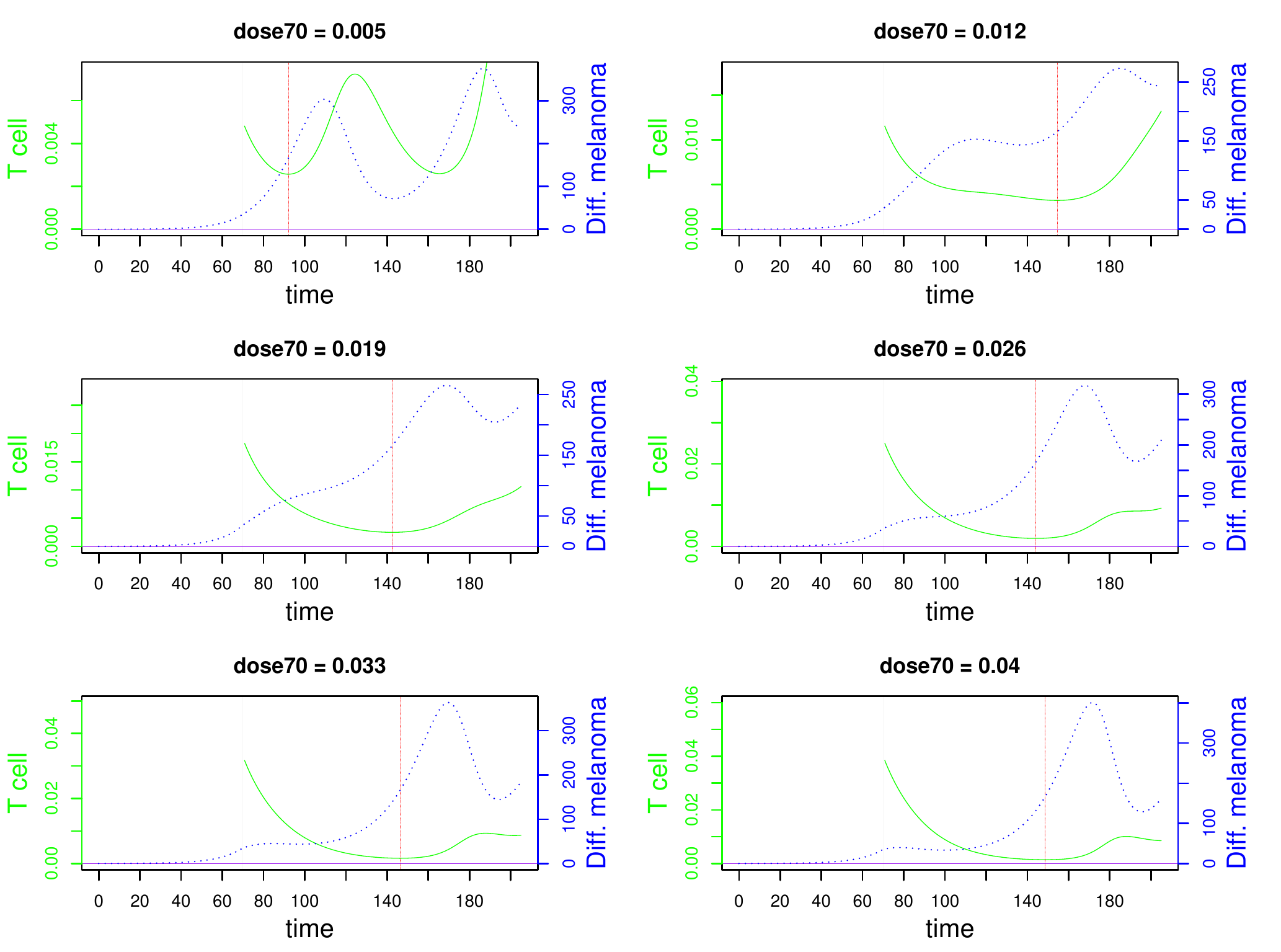}
	\caption{\scriptsize{T cell (solid line in green) and differentiated melanoma cell (dotted line in blue) along time for different doses in ACT group. $d_{MT} =  \text{q}_{50}(t_{T_i})$, $d_T =  \text{q}_{95}(d_{T_i})$,  $l_{A}^{\text{prod}} =  \text{q}_{50}(l_{A_i}^{\text{prod}})$, $n_{M_0} =  \text{q}_{50}(n_{M_{0_i}})$. Vertical line:   treatment start,   vertical line in red indicates the value of $g(d_{70})$.}}
	\label{Tcell_VS_DM_tT_50_dT_95_Lw_50_M0_50_pour_doses_croissantes}
\end{figure}

\end{document}